\documentclass[a4paper,fleqn]{cas-dc}

\usepackage[numbers,sort&compress]{natbib}
\usepackage{hyperref}
\usepackage{xurl}

\begin{document}
\let\WriteBookmarks\relax
\def\floatpagepagefraction{1}
\def\textpagefraction{.001}

\shorttitle{Thick-target Yield of $^{65}$Cu($\alpha,n)^{68}$Ga Near Threshold}    

\shortauthors{Z. Meisel, G. Hamad, K. Brandenburg et al.}  

\title [mode = title]{Thick-target Yield of $^{65}$Cu($\alpha,n)^{68}$Ga Near Threshold and Implications for Nuclear Medicine, Deep Underground Detector Backgrounds, and Nucleosynthesis}  



%

\author[1]{Z. Meisel}[orcid=0000-0002-8403-8879]
\affiliation[1]{organization={Department of Engineering Physics, Air Force Institute of Technology},
            city={Wright-Patterson AFB},
            postcode={45433}, 
            state={Ohio},
            country={U.S.A.}}
\cormark[1]
\ead{zachary.meisel@us.af.mil}

\author[2]{G. Hamad}[orcid=0009-0009-8054-1483]
\cormark[2]
\ead{XSN3YC@uvahealth.org}
\affiliation[2]{organization={ University of Virginia School of Medicine},
            city={Charlottesville},
            postcode={22903}, 
            state={Virginia},
            country={U.S.A.}}

\author[3]{K. Brandenburg}
\affiliation[3]{organization={Institute of Nuclear \& Particle Physics, Department of Physics \& Astronomy, Ohio University},
            city={Athens},
            postcode={45701}, 
            state={Ohio},
            country={U.S.A.}}

\author[3]{C.R. Brune}[orcid=0000-0003-3696-3311]
\author[3]{D.E. Carter}
\author[3]{J. Derkin}
\author[3]{Y. Jones-Alberty}[orcid=0000-0002-0131-2204]
\author[3]{T.N. Massey}[orcid=0000-0002-5578-5357]
\author[3]{D. Soltesz}
\author[3]{S.K. Subedi}[orcid=0000-0001-5898-1989]
\author[3]{A.V. Voinov}

\cortext[1]{Corresponding author}



\begin{abstract}
We report thick target yields of $^{65}{\rm Cu}(\alpha,n)^{68}{\rm Ga}$ from $\alpha$ energies $E_{\alpha}$$=$6.5--8~MeV measured via neutron-counting at the Edwards Accelerator Laboratory at Ohio University. Our results agree with prior results based on single-foil irradiations, while disagreeing with thick-target yields inferred from stacked-foil irradiations by up to an order of magnitude. Based on comparisons of our and prior measurements to Hauser-Feshbach calculations, we estimate the thick-target yield for $^{68}{\rm Ga}$ from $E_{\alpha}$=14~MeV impinging on 100\%-enriched copper ($^{65}{\rm Cu}$) is 161$^{+5}_{-24}$~MBq/$\mu$Ah (for He$^{++}$), which is in agreement with the current IAEA estimate of 156~MBq/$\mu$Ah. At $E_{\alpha}$$=$8~MeV, where radioimpurity-free production of $^{68}{\rm Ga}$ could be realized with a natural copper target, our yield estimate is nearly 4$\times$ lower than the IAEA estimate. We find that the $^{65}{\rm Cu}(\alpha,n)$ cross section estimates presently employed for neutrino and dark matter detector backgrounds and the $^{65}{\rm Cu}(\alpha,n)$ astrophysical reaction rate recently recommended for nucleosynthesis calculations are in agreement with our measurement results.

\end{abstract}

\begin{graphicalabstract}
\includegraphics{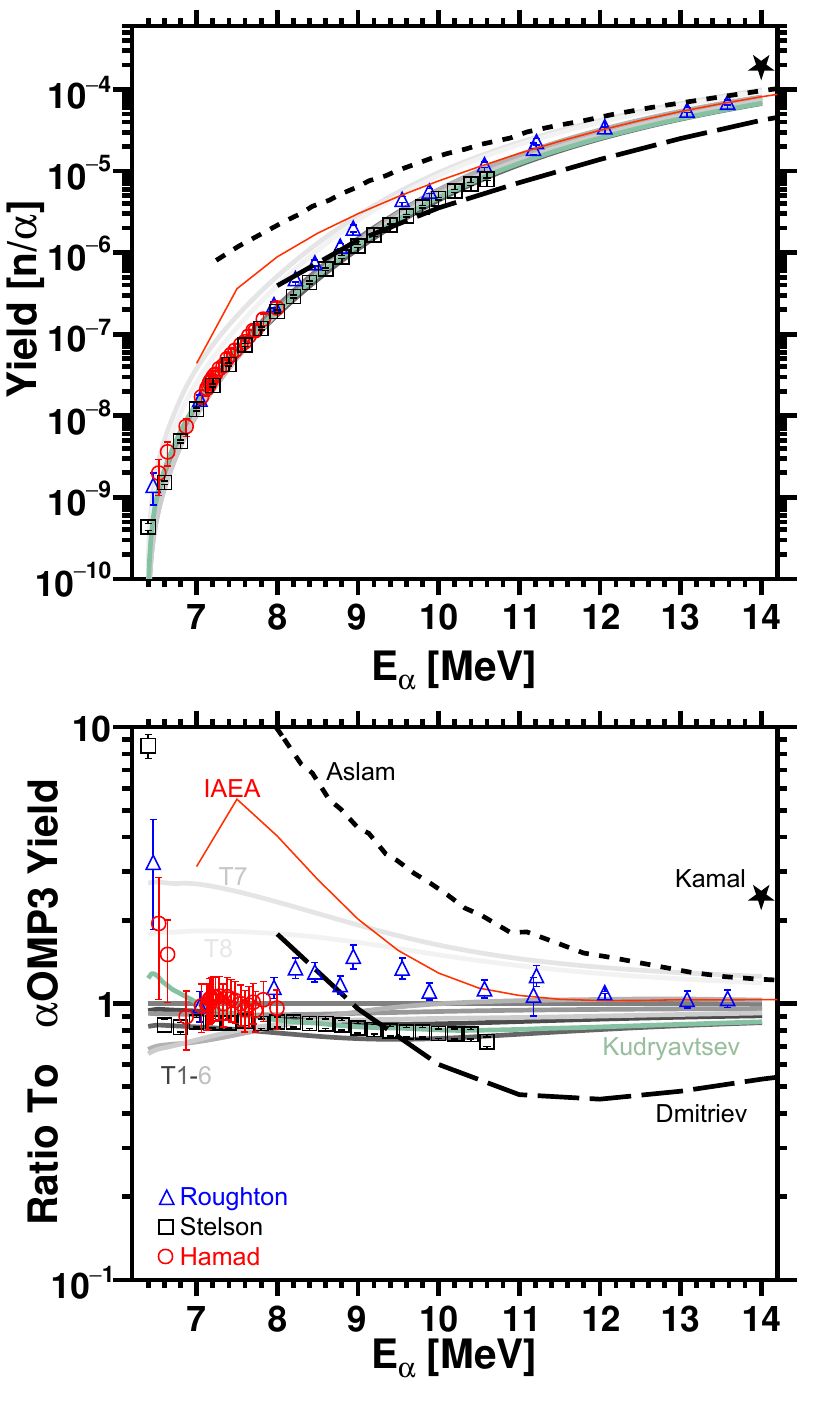}
\end{graphicalabstract}

\begin{highlights}
\item New thick-target yields reported for $^{65}{\rm Cu}(\alpha,n)^{68}{\rm Ga}$ from $E_{\alpha}=6.5-8$~MeV based on neutron counting.
\item Yields agree with prior single-foil activation and neutron counting measurements, and are substantially smaller than estimates based on stacked-foil activation experiments.
\item Based on comparisons of Hauser-Feshbach calculations to this and previous measurements, we estimate the thick-target yield for $^{68}{\rm Ga}$ from $E_{\alpha}=14$~MeV $\alpha$-particles impinging on enriched copper ($^{65}{\rm Cu}$) is 161$^{+5}_{-24}$~MBq/$\mu$Ah (for He$^{++}$), which is in agreement with the current IAEA estimate of 156~MBq/$\mu$Ah. Our yield estimate is nearly 4$\times$ lower than the IAEA estimate for $E_{\alpha}$$=$8~MeV, where radioimpurity-free production of $^{68}{\rm Ga}$ could be achieved with a natural copper target.
\item $^{65}{\rm Cu}(\alpha,n)$ cross section estimates presently employed for neutrino and dark matter detector backgrounds agree with the present results.
\item The astrophysical reaction rate corresponding to our recommended cross section is in agreement with rates calculated using a state-of-the-art $\alpha$-optical potential employed in nucleosynthesis calculations.

\end{highlights}



\maketitle

\section{Introduction}\label{sec:intro}
The nuclear reaction $^{65}{\rm Cu}(\alpha,n)^{68}{\rm Ga}$ is of interest for a wide variety of applications, including nuclear medicine, neutrino and dark matter detection, and nuclear astrophysics~\citep{Qaim2017,Tarkanyi2019,Kudryavtsev2008,Roughton1983}. In nuclear medicine, this reaction is an avenue for production of the medical isotope $^{68}{\rm Ga}$. In neutrino and dark matter detection, the presence of $\alpha$-emitting actinides and copper components in detection equipment results in a background of massive neutral particles (neutrons) in deep underground detectors dedicated to detecting other neutral particles, namely neutrinos and dark matter~\citep{Westerdale2017}. In helium-rich astrophysical environments, $^{65}{\rm Cu}(\alpha,n)$ is one reaction (of many) involved in nucleosynthesis of heavier elements. 

$^{68}{\rm Ga}$ medical isotope production has been of particular interest for the majority of past $^{65}{\rm Cu}(\alpha,n)$ measurements. $^{68}{\rm Ga}$ is desirable for medical imaging and diagnostics as it is a positron emitter with a convenient half-life~\citep{Baum2012,Banerjee2013}. When a high yield is desired, one promising route for $^{68}{\rm Ga}$ production is $^{65}{\rm Cu}(\alpha,n)$, where maintaining $\alpha$ beam energies $E_{\alpha}$$<$14~MeV avoids production of undesirable radioimpurities~\citep{Aslam2021}. This energy regime overlaps with $E_{\alpha}$$\lesssim$8~MeV of interest for $(\alpha,n)$ backgrounds due to $\alpha$-emitting actinides~\citep{Brandenburg2023a} and the near-threshold energies typically of interest for astrophysical reaction rates~\citep{Rauscher2010}. $E_{\alpha}$$=$8~MeV is meanwhile of potential interest for medical isotope production with natural copper targets, as it avoids all radiosotope contamination~\citep{Scelecsenyi2012}.

A number of $^{65}{\rm Cu}(\alpha,n)^{68}{\rm Ga}$ cross section and thick-target yield measurements have been performed to-date covering $E_{\alpha}$$<$14~MeV, approaching the nuclear reaction threshold (6.139~MeV~\citep{Wang2021})~\citep{Bryant1963,Stelson1964,Roughton1983,Dmitriev1986,Bhardwaj1988,Levkovskij1991,Bonesso1991,Singh1994,Scelecsenyi2012,Hermanne2015}. All except for Refs.~\citep{Stelson1964,Roughton1983} performed their measurements using the stacked-foil activation method with an $\alpha$-beam from a cyclotron. These measurements generally start with large $E_{\alpha}$, such that the low-energy yields are derived from foils located deep within the stack. Several uncertainties in $E_{\alpha}$ compound in such measurements~\citep{Scelecsenyi2012}. The incident beam energy generally begins with a hundreds-of-keV uncertainty when produced by a cyclotron and is then degraded by energy loss through the foil-stack. Uncertainties in energy loss and energy and angular straggling, along with uncertainties in initial beam alignment, can readily compound to result in MeV-level errors.

Specific concerns with existing $^{65}{\rm Cu}(\alpha,n)$ data include the following. The cross section data from \citet{Bryant1963} are too-sparse near threshold to enable a thick-target yield calculation. \citet{Dmitriev1986} report data from previous measurements without specifying any experimental details.  \citet{Bhardwaj1988} report a single data-point with $E_{\alpha}$$<14$~MeV with a large energy uncertainty. \citet{Levkovskij1991} report (unphysical) non-zero yields below threshold and have a large energy uncertainty. \citet{Bonesso1991} report cross sections systematically much larger than all other datasets at which there is a common measurement energy, leading to their exclusion from an IAEA evaluation of medical isotope production cross sections (along with \citet{Levkovskij1991} for the aforementioned reasons)~\citep{Tarkanyi2019}. \citet{Singh1994} report a single data-point within $E_{\alpha}$$<$14~MeV. The cross sections reported by \citet{Scelecsenyi2012} and \citet{Hermanne2015} each have very large energy uncertainties, where the latter reports non-zero cross sections below threshold (due to the large energy uncertainty).

Various subsets of these data have been employed to estimate the medical isotope production yield for $^{65}{\rm Cu}(\alpha,n)^{68}{\rm Ga}$. In general this involves obtaining a single-valued function for the nuclear reaction cross section and the stopping power of $\alpha$-particles in copper. As the latter is relatively well known~\citep{Paul2005,Montanari2017}, the former is of particular concern. Smooth approximations for the $^{65}{\rm Cu}(\alpha,n)^{68}{\rm Ga}$ cross section for medical isotope production have been obtained in the following ways. The IAEA evaluation of \citet{Tarkanyi2019} performed a fit via a Pad\'{e} approximation to the stacked-foil activation data of Refs.~\citep{Bryant1963,Bhardwaj1988,Singh1994,Scelecsenyi2012,Hermanne2015} (for $E_{\alpha}$$<$14~MeV). \citet{Aslam2021} fit the data of Refs.~\citep{Bryant1963,Stelson1964,Bhardwaj1988,Bonesso1991,Levkovskij1991,Singh1994,Scelecsenyi2012,Hermanne2015} (for $E_{\alpha}$$<$14~MeV) by normalizing Hauser-Feshbach~\citep{Hauser1952} calculation results with third-order polynomials derived from fits to residuals from the Hauser-Feshbach calculations. \citet{Dmitriev1986} linearly interpolated the cross section data contained within their report. \citet{Kamal2024} performed Hauser-Feshbach calculations and used the data of Ref.~\citep{Hermanne2015} as a benchmark by which to judge their results. Notably, none of the prior estimates leverage the single-foil irradiation data of \citet{Roughton1983} and the one estimate that leverages the single-foil irradiation data of \citet{Stelson1964} also includes a number of stacked-foil datasets (which may be problematic near threshold), thus minimizing the influence of this dataset.

The corresponding medical isotope production cross sections from these analyses vary considerably. At $E_{\alpha}$$=$14 MeV for a 100\%-enriched $^{65}{\rm Cu}$ target, the extreme estimates are 80 MBq/$\mu$Ah \citep{Dmitriev1986} and $\sim$370 MBq/$\mu$Ah \citep{Kamal2024}\footnote{We infer this yield from the statement in \citet{Kamal2024} that their thick target yield is nearly twice that of \citet{Aslam2021}, as \citet{Kamal2024} only state a thick-target yield for $E_{\alpha}$$>$40~MeV, where maximal yield is obtained. Therefore, we take the $E_{\alpha}$$=$14~MeV estimate from \citet{Aslam2021} and multiply it by two. See Sec.~\ref{sec:yield} for further discussion.}. Intermediate estimates are 156 MBq/$\mu$Ah from the IAEA {\it Charged-particle cross section database for medical radioisotope production}\footnote{\url{https://www-nds.iaea.org/medical/}}~\citep{Tarkanyi2019}  and 183 MBq/$\mu$Ah \citep{Aslam2021}. Meanwhile, \citet{Dmitriev1986} report an uncertainty of 10-16\%, where the yield uncertainty implied from the evaluated cross section uncertainty of \citet{Tarkanyi2019} is of similar order when also considering typical stopping-power uncertainties. As such, current estimates of the $^{65}{\rm Cu}(\alpha,n)^{68}{\rm Ga}$ medical isotope production yield are significantly discrepant from one another.

In an effort to resolve these discrepancies, we performed thick-target yield measurements of $^{65}{\rm Cu}(\alpha,n)^{68}{\rm Ga}$ and compared our results and earlier single-foil irradiation results to Hauser-Feshbach calculations. Section~\ref{sec:measurements} describes our experiment and analysis to obtain the thick-target yield. Section~\ref{sec:yield} presents the yield in comparison to previous results and Hauser-Feshbach calculations, in the context of $^{68}{\rm Ga}$ medical isotope production. Section~\ref{sec:csrate} describes the inferred $^{65}{\rm Cu}(\alpha,n)^{68}{\rm Ga}$ cross section and implications for neutrino and dark matter detector backgrounds, as well as for astrophysical nucleosynthesis calculations. Section~\ref{sec:conclusions} provides conclusions.

\section{Measurement and Analysis}\label{sec:measurements}

\begin{figure}
  \centering
\includegraphics[width=0.9\columnwidth]{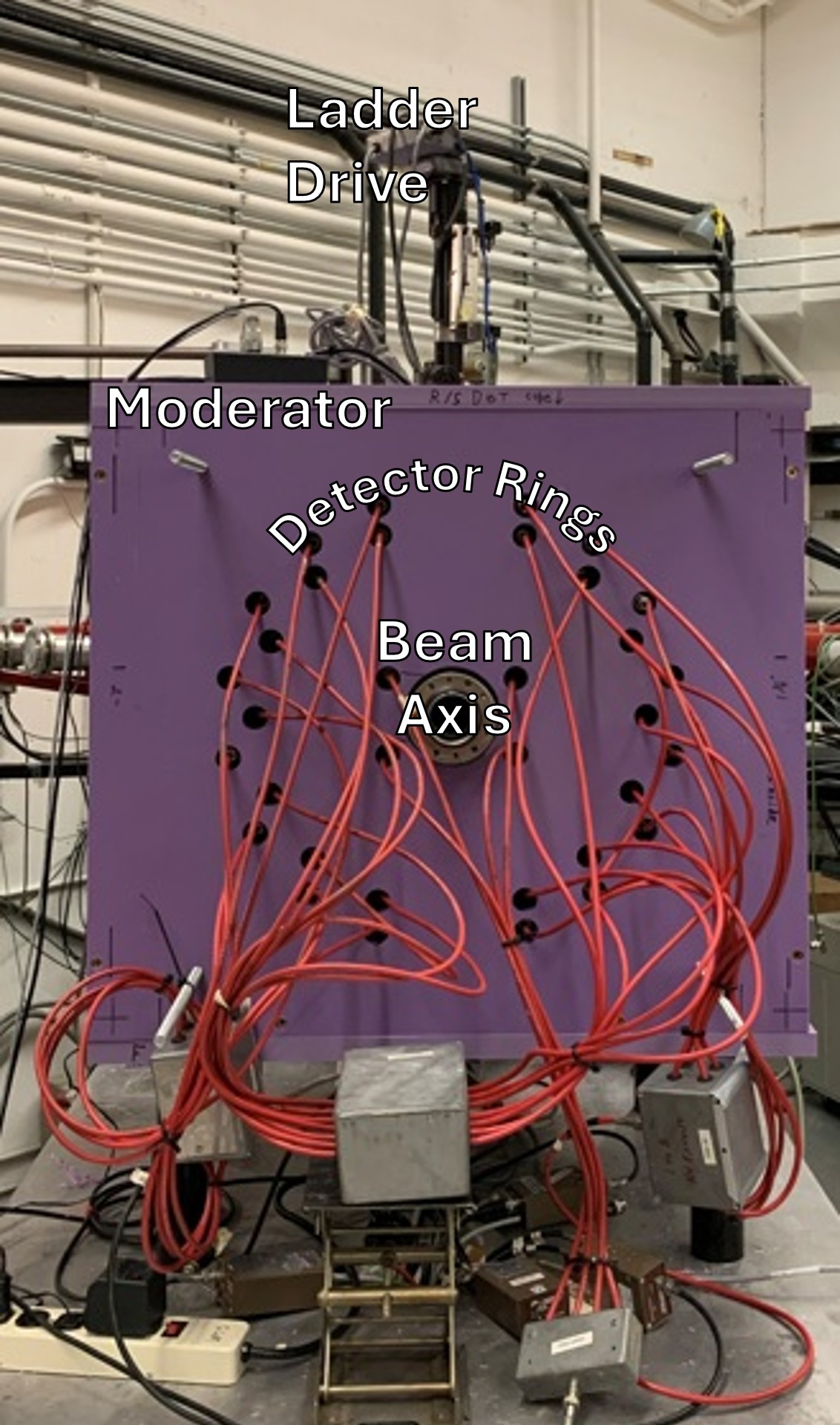}
    \caption{The HeBGB neutron detector. A purple borated polyethylene box surrounds the polyethylene moderator, in which the $^{3}$He and BF$_{3}$ proportional counters are embedded in a concentric ring configuration about the beam axis. A ladder drive enables quick change between targets located along the beam axis.}\label{fig:HeBGB}
\end{figure}

Thick-target yield measurements were performed between $E_{\alpha}$=$6.533-7.996$~MeV at the Edwards Accelerator Laboratory~\citep{Meisel2017} using the $^{3}{\rm He}$BF$_{3}$ Giant Barrel (HeBGB) experiment set-up described in Refs.~\citep{Brandenburg2022,Brandenburg2023b} and shown in Fig.~\ref{fig:HeBGB}. 

Helium ions were produced using an Alphatross ion source and accelerated with a 4.5~MV T-type tandem Pelletron, departing the carbon stripper foil located within the accelerator terminal as He$^{++}$. The beam was bent toward the 65$^{\circ}$-right beamline of the large target room via the energy-calibrated 90$^{\circ}$ analyzing magnet. The beam energy is calibrated with 0.1\% uncertainty using an NMR magnetometer on the field of the analyzing magnet. The beam energy resolution and tune-dependent energy variations (for a given NMR setting) are limited to being <0.2\% by the 0.152-cm gap at the magnet exit. The beam passed through an 0.5-cm-diameter gold-plated collimator located approximately 48~cm upstream of the target location prior to entering the beam-pipe within HeBGB. This and all-other gold-plating referred to for the present measurement consisted of 99.95\% purity 0.05-mm-thick gold foil from Avantor. The beam-current impinging on the set-up was integrated with 1\% uncertainty for the beam-pipe within HeBGB, the target ladder, and, for collimator measurements, the downstream Faraday cup. Typical on-target beam currents were $\sim$10-50~nA, corresponding to beam intensities of $I_{\alpha}$$\sim$$10^{11}$~pps.

\begin{figure}
  \centering
\includegraphics[width=0.5\columnwidth]{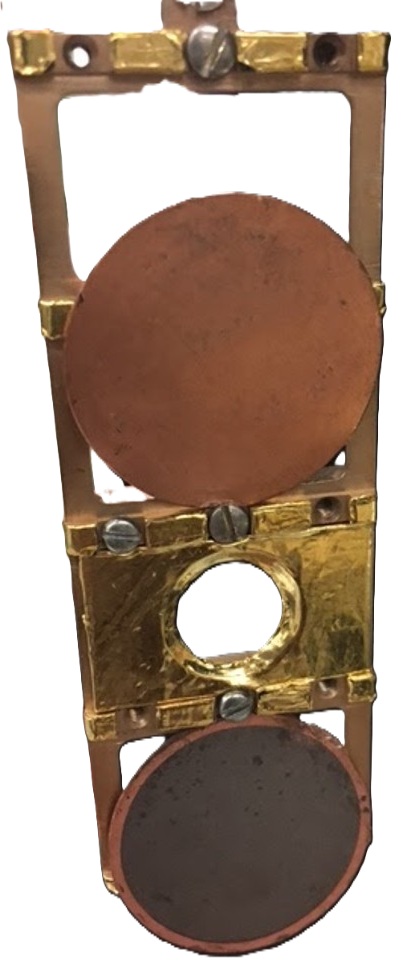}
    \caption{The gold-plated target ladder with copper, collimator, and $^{13}$C targets, from top to bottom.}\label{fig:Ladder}
\end{figure}

The target consisted of the gold-plated ladder shown in Fig.~\ref{fig:Ladder}, mounted at the end of the drive labeled in Fig.~\ref{fig:HeBGB}. During tuning, the drive was adjusted such that the centrally-positioned gold-plated collimator was located along the beam axis. The beam was tuned in order to achieve a 0~nA current reading on the ladder, while maximizing current on the downstream Faraday cup. The drive was then adjusted such that the beam was centered on the upper target, an 0.8-mm-thick disk of oxygen-free copper with natural abundance manufactured by the Duke machine shop. To correct for the neutron background from carbon contamination, as described shortly, measurements (described in full in Ref.~\citep{Brandenburg2023a}) were also performed at each energy with the drive adjusted such that the beam was centered on the bottom target, $^{13}{\rm C}$ enriched to $>$99\% evaporated onto a copper substrate by Arizona Carbon Foil. Beam-on-target times ranged from approximately $t_{\rm meas.}$$=$0.5--3~minutes for each target+$E_{\alpha}$ combination. 

$N_{\rm detected}$ neutrons were detected using $^{3}$He and BF$_{3}$ proportional counters, from Baker Hughes and Reuter-Stokes, respectively,  located within concentric rings about the beam axis, embedded within a block of ultra-high-molecular-weight polyethylene from Trident Plastics, as shown in Fig.~\ref{fig:HeBGB} and described in Ref.~\citep{Brandenburg2022}. The polyethylene moderator was surrounded by a 2.54-cm-thick box of 5\% borated high-density polyethylene from Professional Plastics to reduce external neutron background. The HeBGB neutron detection efficiency was $\eta$$=$7.5$\pm$1.2\%, as determined in Ref.~\citep{Brandenburg2022}, where HeBGB provides nearly 4$\pi$ solid-angle coverage. The angular distribution of neutrons from $^{65}{\rm Cu}(\alpha,n)$ was assumed to be isotropic, consistent with expectations for $(\alpha,n)$ reactions near-threshold on medium-mass nuclides~\citep{Douglas1959,Grimes1974}, and so angular-distribution corrections to the efficiency were not performed. A 60~Hz pulser was adjusted to provide a signal outside of the neutron spectrum in order to monitor the live-fraction $f_{\rm live}$ of the data acquisition. As count rates were $\lesssim$200~n/s at all measurement energies, $f_{\rm live}$$\approx$1. 

\begin{figure}
  \centering
\includegraphics[width=1.0\columnwidth]{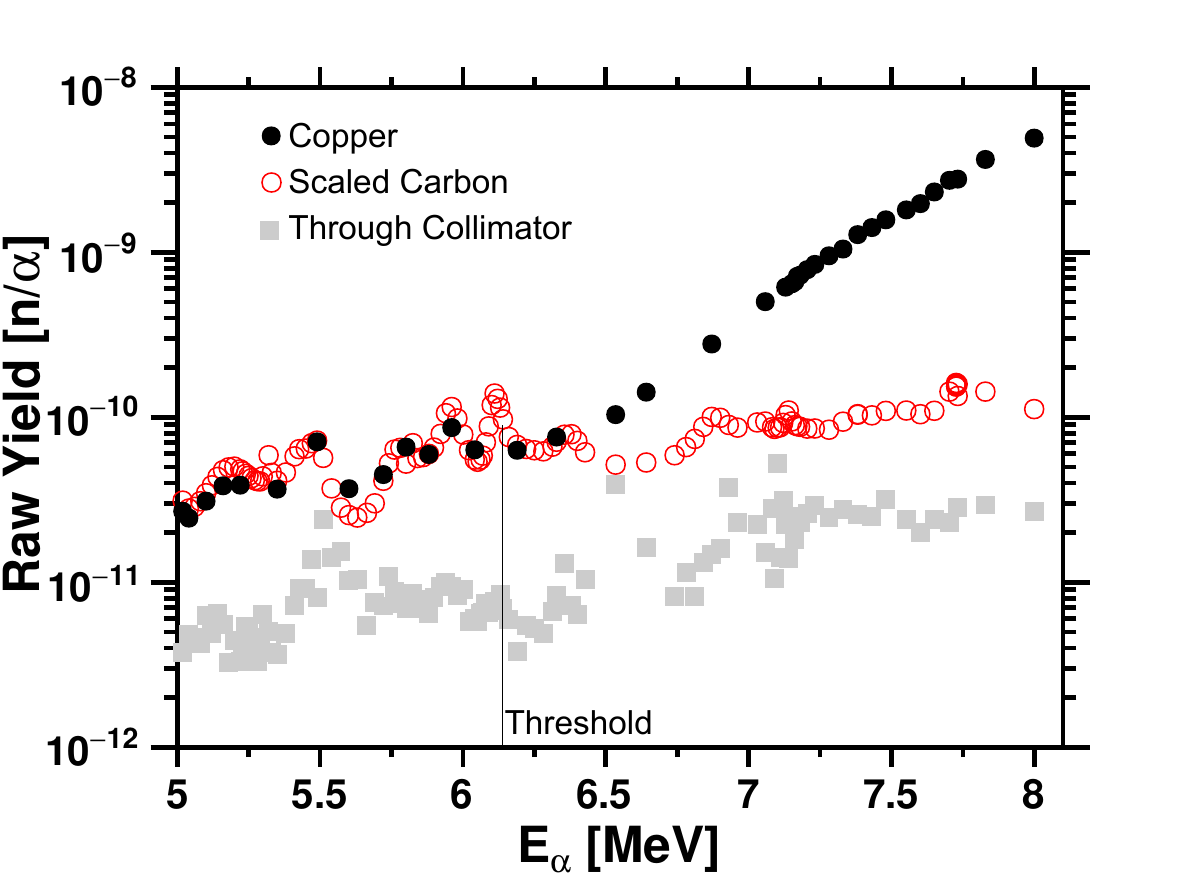}
    \caption{Raw yield measured with the copper target (solid black circles) or collimator (solid gray squares) located at the beam axis. The measured $^{13}$C target yield has been scaled by an arbitrary factor (open red circles) to overlap the low-$E_{\alpha}$ copper data. The vertical line indicates the $^{65}{\rm Cu}(\alpha,n)$ threshold.}\label{fig:BackgroundYield}
\end{figure}

The ambient neutron background was approximately 0.4~n/s. However, it was apparent that beam-induced background was more significant, given the non-zero yields we obtained for $^{65}{\rm Cu}(\alpha,n)$ below threshold following ambient background subtraction. Similarly, subtraction of the yield measured when the beam was tuned through the target-ladder collimator was insufficient to remove the below-threshold non-zero yield. We observed that the below-threshold yield on the copper target closely matched the trend measured for the $^{13}$C target, once the latter was scaled down, as shown in Fig.~\ref{fig:BackgroundYield}. We therefore attribute our neutron background $N_{\rm background}$ to $^{13}{\rm}$C-contamination on the copper target and apply this scaled-down yield as a background subtraction, to which we assign an uncertainty of 6$\times$10$^{-10}$~n/$\alpha$ (after accounting for detection efficiency). We do not consider the possibility of carbon build-up over time modifying the contamination level~\citep{Blondiaux1984}, since the target accumulated relatively small amount of beam charge at relatively low intensity and measurements were performed over the course of roughly one day. We note that the trend in the yield for the $^{13}$C target lacks the peak observed in the yield around $E_{\alpha}$$\approx$6.5~MeV when the beam was tuned through the collimator. As it is unknown whether the neutrons from the through-collimator measurements originate from the downstream Faraday cup or upstream beamline components, it is unknown whether this background peak would be present in the copper-target data. As such, it is possible that our $^{65}{\rm Cu}(\alpha,n)$ yields for the $E_{\alpha}$$\approx$6.5 and 6.6~MeV data points are somewhat overestimated.

The measured thick-target yield was calculated as \begin{equation}
Y_{\rm meas.}=\frac{(N_{\rm detected}-N_{\rm background})}{I_{\alpha}t_{\rm meas.}\eta f_{\rm live}},
\end{equation}
where we note that the conversion from beam current to beam intensity accounts for the doubly-ionized nature of the helium beam. The thick target yield uncertainty sources were 1--4\% statistical, 1\% beam current integration, 16\% HeBGB efficiency, and $\pm$6$\times10^{-10}$~n/$\alpha$ carbon background subtraction, all of which were summed in quadrature to arrive at a total yield uncertainty. Though our measurements were performed with a natural copper target, the thick-target yield of a pure $^{65}{\rm Cu}$ target is of interest. Therefore we divide our measured yields by the natural abundance fraction of $^{65}{\rm Cu}$, 0.3085~\citep{Berglund2011}.

\section{Thick-Target Yield and Nuclear Medicine Implications}\label{sec:yield}

\begin{table}
\caption{Thick-target yield for $^{65}{\rm Cu}(\alpha,n)$ determined in this work.}\label{tab:yield}
\begin{tabular*}{\tblwidth}{@{}LLLL@{}}
\toprule
  &  \\ 
$E_{\alpha}$ [MeV]	&	$dE_{\alpha}$ [MeV]	&	Yield [n/$\alpha$]	&	dYield [n/$\alpha$]	\\
\midrule
6.533	&	0.014	&	1.98$\times 10^{-9}$	&	9.28$\times 10^{-10}$	\\
6.642	&	0.014	&	3.61$\times 10^{-9}$	&	1.19$\times 10^{-9}$	\\
6.866	&	0.015	&	7.39$\times 10^{-9}$	&	1.79$\times 10^{-9}$	\\
7.057	&	0.015	&	1.71$\times 10^{-8}$	&	3.38$\times 10^{-9}$	\\
7.128	&	0.015	&	2.17$\times 10^{-8}$	&	4.12$\times 10^{-9}$	\\
7.150	&	0.015	&	2.35$\times 10^{-8}$	&	4.43$\times 10^{-9}$	\\
7.161	&	0.015	&	2.54$\times 10^{-8}$	&	4.74$\times 10^{-9}$	\\
7.171	&	0.015	&	2.64$\times 10^{-8}$	&	4.91$\times 10^{-9}$	\\
7.182	&	0.015	&	2.75$\times 10^{-8}$	&	5.08$\times 10^{-9}$	\\
7.204	&	0.015	&	2.98$\times 10^{-8}$	&	5.46$\times 10^{-9}$	\\
7.231	&	0.015	&	3.23$\times 10^{-8}$	&	5.85$\times 10^{-9}$	\\
7.281	&	0.016	&	3.78$\times 10^{-8}$	&	6.75$\times 10^{-9}$	\\
7.330	&	0.016	&	4.10$\times 10^{-8}$	&	7.27$\times 10^{-9}$	\\
7.379	&	0.016	&	5.00$\times 10^{-8}$	&	8.75$\times 10^{-9}$	\\
7.428	&	0.016	&	5.63$\times 10^{-8}$	&	9.79$\times 10^{-9}$	\\
7.477	&	0.016	&	6.35$\times 10^{-8}$	&	1.10$\times 10^{-8}$	\\
7.548	&	0.016	&	7.45$\times 10^{-8}$	&	1.27$\times 10^{-8}$	\\
7.597	&	0.016	&	8.06$\times 10^{-8}$	&	1.37$\times 10^{-8}$	\\
7.646	&	0.016	&	9.46$\times 10^{-8}$	&	1.60$\times 10^{-8}$	\\
7.701	&	0.016	&	1.11$\times 10^{-7}$	&	1.86$\times 10^{-8}$	\\
7.728	&	0.016	&	1.11$\times 10^{-7}$	&	1.85$\times 10^{-8}$	\\
7.826	&	0.017	&	1.53$\times 10^{-7}$	&	2.52$\times 10^{-8}$	\\
7.996	&	0.017	&	2.10$\times 10^{-7}$	&	3.44$\times 10^{-8}$	\\

\bottomrule
\end{tabular*}
\end{table}

\begin{figure}
  \centering
\includegraphics[width=1\columnwidth]{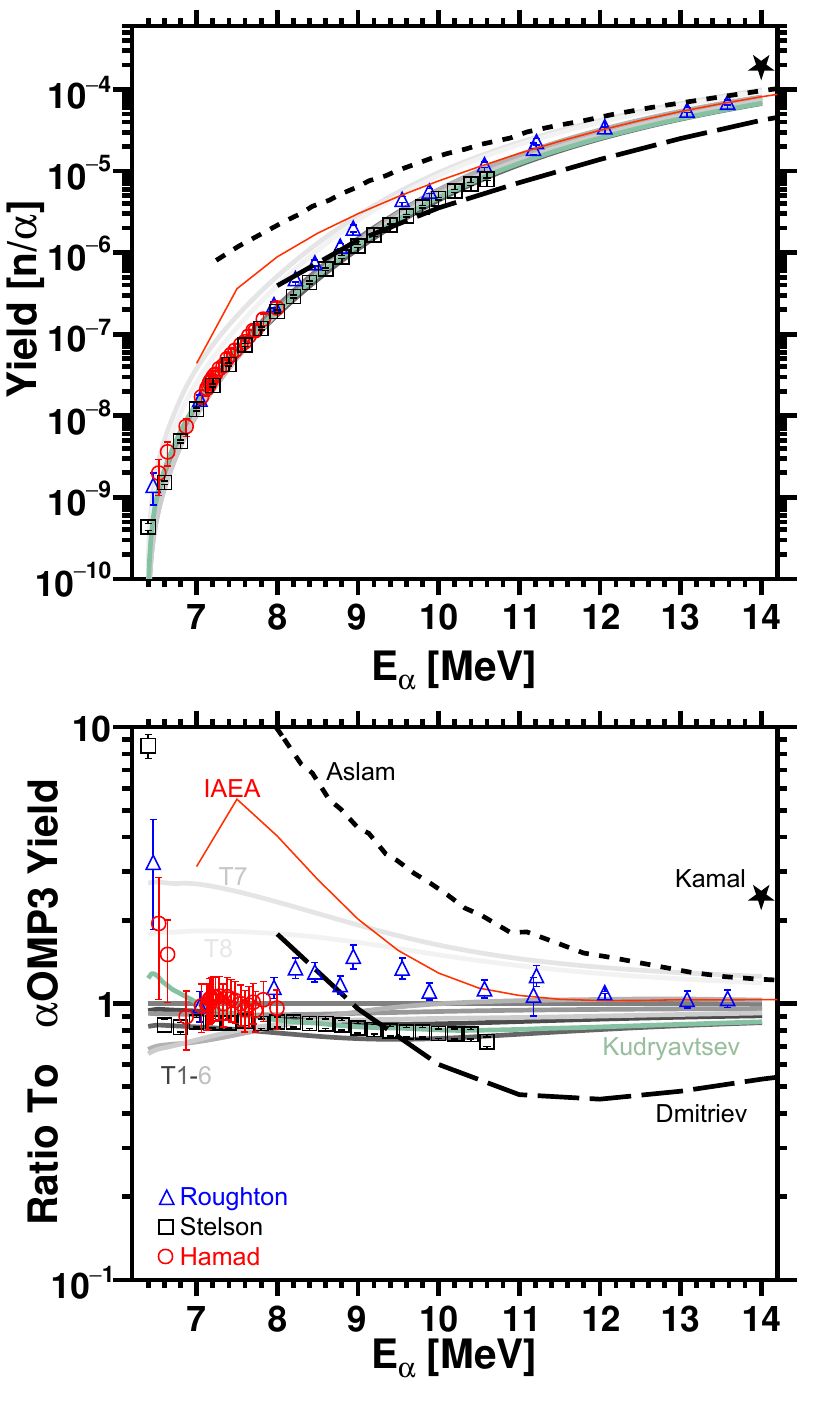}
    \caption{(upper panel) Thick-target yield for $^{65}{\rm Cu}(\alpha,n)$ from this work's (Hamad) and prior experimental data~\citep{Stelson1964,Roughton1983} along with several evaluations~\citep{Dmitriev1986,Kudryavtsev2008,Tarkanyi2019,Aslam2021,Kamal2024} and the Hauser-Feshbach calculations performed with $\alpha$OMP 1--8 (T1--T8). The lower-panel shows the ratio of the upper panel data to the $\alpha$OMP model 3 yield for ease of comparison.}\label{fig:Yield}
\end{figure}

Our thick-target yield for $^{65}{\rm Cu}(\alpha,n)^{68}{\rm Ga}$ is reported in Tab.~\ref{tab:yield}. In Fig.~\ref{fig:Yield}, we compare our results to existing experimental thick-target yield data~\citep{Stelson1964,Roughton1983} and the myriad of evaluations and estimates discussed in Section~\ref{sec:intro}. Within our covered energy range, we agree with both the activation data of \citet{Roughton1983} and the neutron-counting data of \citet{Stelson1964}, though these two are in tension within their stated uncertainties. Only \citet{Tarkanyi2019} provide a yield estimate for medical isotope production within our measured energy range, which is approximately 3--6$\times$ our measured yield. This is likely due to the fact that the IAEA analysis focused on higher $E_{\alpha}$ and fit to stacked-foil activation results, which struggle to provide accurate results at low energies, as described in Section~\ref{sec:intro}.

In order to compare to medical isotope production estimates up to $E_{\alpha}$$=$14~MeV, we performed Hauser-Feshbach calculations of $^{65}{\rm Cu}(\alpha,n)$ with {\tt Talys} version 2.0~\citep{Koning2023} for a variety of $\alpha$-optical model potentials ($\alpha$OMP) and calculated thick-target yields assuming the helium stopping power in copper from {\tt SRIM2013}~\citep{Ziegler2010}. Hauser-Feshbach calculations of $(\alpha,n)$ reactions on medium-mass nuclides are generally considered to accurately predict experimental cross-sections within an order of magnitude~\citep{Mohr2015}, where the spread in predictions near threshold is predominantly due to the $\alpha$OMP~\citep{Rauscher2012}. For the present case, we find $<$10\% impact on the calculated $^{65}{\rm Cu}(\alpha,n)$ cross section when exploring nuclear level-density and neutron and proton optical potential model options available in {\tt Talys}. Therefore, once anchored by experimental data, Hauser-Feshbach calculations are expected to provide an accurate representation of the energy trend in the $^{65}{\rm Cu}(\alpha,n)$ cross section (i.e. excitation function) and therefore extrapolation from our low $E_{\alpha}$ dataset up to $E_{\alpha}$$=$14~MeV.

{\tt Talys} calculations were performed for $\alpha$OMP models 1 through 8~\citep{Koning2023}, using default model parameters otherwise. We calculated the thick-target yield from the Hauser-Feshbach cross section $\sigma_{\rm HF}$ and stopping-power $\epsilon(E)$ via
\begin{equation}
    Y_{\rm calc.}=\int^{E_{\alpha}}_{0} \frac{\sigma_{\rm HF}(E)}{\epsilon(E)}dE,
\end{equation}
resulting in the yield curves shown in Fig.~\ref{fig:Yield}. For ease of comparison, the lower-panel shows the ratio of all thick-target yields discussed in this work to the yield obtained with $\alpha$OMP model 3, as this model selection best reproduces the experimental data. We note however that yields calculated using $\alpha$OMP models 1--6~\citep{Koning2023,McFadden1966,Demetriou2002,Avrigeanu2014} are all within one standard deviation of our $Y_{\rm meas.}$ and are bounded by the thick-target yields of \citet{Stelson1964} and \citet{Roughton1983}. Meanwhile, yields calculated using $\alpha$OMP models 7~\citep{Nolte1987} and 8~\citep{Avrigeanu1994} are clearly discrepant with the experimental data. In principle, these $Y_{\rm calc.}$ have some uncertainty contribution from the choice of $\epsilon(E)$. In practice, the {\tt SRIM2013} $\epsilon(E)$ for $E_{\alpha}$$=$6--14~MeV helium within copper is within 1\% of all experimental data\footnote{\url{https://nds.iaea.org/stopping/tuples/He-Cu}}~\citep{Andersen1977,Ishiwari1975}, with the exception of a single data-point~\citep{Ishiwari1977} that is discrepant by 6\%, and within 1\% of other commonly employed $\epsilon(E)$ theory estimates~\citep{Berger2017,Bivort2022}\footnote{We note that the Bethe-Bloch approach of \citet{Leo1994} that is employed by {\tt Talys} for medical isotope production calculations is larger by 7\%.}. Thus the $\epsilon(E)$ uncertainty is negligible in the present context.

At the lowest $E_{\alpha}$ in Fig.~\ref{fig:Yield}, each measurement appears to diverge from the Talys predictions using built-in $\alpha$OMP, hinting at larger-than-predicted cross sections near threshold. Hauser-Feshbach calculations of $\alpha$-induced reactions near threshold are known to be extremely sensitive to the the tail of the adopted $\alpha$OMP~\citep{Mohr2015}. Thus, it could be that in this case a model tailored to address near-threshold energies (e.g. \citep{Mohr2020}) would be more appropriate for this lowest-energy region. As these lowest-energy data points are not the main focus of the present work, we do not perform such a comparison.

For additional comparisons, we must convert between the thick-target yield in units of $n/\alpha$ to units of MBq/$\mu$Ah, where the latter is more standard in the language of medical isotope production~\citep{iaea2001}. We denote the latter yield (often referred to as the physical yield~\citep{Otuka2015}) as $\tilde{Y}$ and note that, though expressed with units of activity, it does not account for radioactive decay of the produced nuclide during production or a cool-down period{\footnote{Yet a third standard is the saturated thick-target yield~\citep{Otuka2015}, which does account for decay of the product during irradiation. For this standard, $t_{\rm irr.}$ of Eqn.~\ref{eqn:Atirr} is set to $\infty$. Accounting for cooling following irradiation is shown in Eqn.~\ref{eqn:Atcool}.}. The conversion is
\begin{equation}
    \tilde{Y}=\frac{\lambda}{q}Y,
    \label{eqn:conversion}
\end{equation}
where $\lambda$ is the decay constant of the produced nuclide and $q$ is the electric charge of the incident beam. We emphasize that the standard for helium beams with $E_{\alpha}$ of several MeV or more is $q$$=$$2e$, as this is the charge-state realized in accelerators producing helium nuclei at these energies. We emphasize this because it appears to have been cause for some confusion in the literature. Namely, when \citet{Aslam2021} re-report the results of \citet{Dmitriev1986}, they quote yields that are smaller by \textonehalf\, from that reported in the original work (when the original work is converted to an enriched $^{65}{\rm Cu}$ abundance). This is presumably under the assumption that \citet{Dmitriev1986} either adopted $q$$=$1$e$ or meant particle microamps ($p\mu$A) instead of electrical microamps ($e\mu$A), though \citet{Dmitriev1986} are explicit that $e\mu$A are their meaning\footnote{We mean $e\mu A$ in the present work as well.} and they note that the relative charge of the bombarding particle is accounted for, i.e. $q$$=$2$e$. Meanwhile \citet{Kamal2024} appear to misreport the result for their own $\tilde{Y}$. Those authors recommend the $\tilde{Y}$ calculated based on $\sigma_{\rm HF}$ from {\tt Talys} using $\alpha$OMP model 3, stating it is 1558~MBq/$\mu$Ah for $E_{\alpha}$ providing maximum yield ($E_{\alpha}>40$~MeV). However, employing Eqn.~\ref{eqn:conversion} for this $\sigma_{\rm HF}$ yields a result of 680~MBq/$\mu$Ah for $E_{\alpha}=40$~MeV. This is in-line with expectations, considering the IAEA yield at this energy is 796~MBq/$\mu$Ah~\citep{Tarkanyi2019} and the fact that the cross section calculation of \citet{Kamal2024} slightly undershoots the \citet{Hermanne2015} dataset that anchors the IAEA cross section fit (See Fig.~6 of Ref.~\citep{Kamal2024}). We speculate that this $\sim$$\times2$ overestimate is due to an issue with $q$ in the $Y$$\rightarrow$$\tilde{Y}$ conversion.

After applying Eqn.~\ref{eqn:conversion} to the $\alpha$OMP 1--6 $Y_{\rm calc.}$ and adopting $\alpha$OMP model 3 as the central estimate, we calculate the thick-target yield for $^{68}{\rm Ga}$ from $E_{\alpha}$$=$14~MeV $\alpha$-particles impinging on 100\%-enriched copper ($^{65}{\rm Cu}$) is 161$^{+5}_{-24}$~MBq/$\mu$Ah (for He$^{++}$). This agrees well with the IAEA estimate 156 MBq/$\mu$Ah \citep{Tarkanyi2019}, is twice the result of \citet{Dmitriev1986}\footnote{If our inference that \citet{Dmitriev1986} used $q$$=$2 is incorrect, then our result would be 4$\times$ that work.}, and is somewhat lower $(12\%)$ than the result of \citet{Aslam2021}. As discussed above, it should agree with \citet{Kamal2024}, as they recommend essentially the same $\sigma_{\rm HF}$.

To avoid the expense and complication of employing 100\%-enriched $^{65}{\rm Cu}$, it may be advantageous to perform $^{65}{\rm Cu}(\alpha,n)^{68}{\rm Ga}$ medical isotope production with natural copper targets. However, this would likely require irradiation at $E_{\alpha}$$=$8~MeV in order to avoid radioisotope contamination from $^{63}{\rm Cu}(\alpha,n)^{66}{\rm Ga}$, whose threshold is 7.923~MeV~\citep{Wang2021}. Using our $Y_{\rm meas.}$ from Tab.~\ref{tab:yield} and assuming He$^{++}$ on natural copper, we estimate a thick-target yield of 0.12$\pm$0.02~MBq/$\mu$Ah. Meanwhile, the IAEA estimate~\citep{Tarkanyi2019} for these conditions is roughly 4$\times$ larger at 0.52~MBq/$\mu$Ah. The \citet{Dmitriev1986} and \citet{Aslam2021} estimates are larger by $\times$2 and $\times$11 at 0.23 and 1.42~MBq/$\mu$Ah, respectively.

We briefly consider whether the yield for $E_{\alpha}$$=$8~MeV on natural copper is large enough to be practically useful. Medical imaging with $^{68}{\rm Ga}$ requires an activity around 180~MBq~\citep{Walker2013}. The activity $A$ of the target immediately following irradiation for a time $t_{\rm irr.}$ is
\begin{equation}
    A(t_{\rm irr.})=I_{\alpha}Y\left(1-e^{-\lambda t_{\rm irr.}}\right)
    \label{eqn:Atirr}
\end{equation}
and after a cooling time $t_{\rm cool}$ following irradiation is
\begin{equation}
    A(t_{\rm cool})=A(t_{\rm irr.})e^{-\lambda t_{\rm cool}}.
    \label{eqn:Atcool}
\end{equation}
Since $\lambda$$=$1.7$\times$$10^{-4}$~s$^{-1}$ for $^{68}{\rm Ga}$~\citep{Mccutchan2012}, $A(t_{\rm irr.})$ reaches 75\% saturation after $t_{\rm irr.}$$\approx$2~h. Chemical separation of radiogallium from copper is possible with high extraction efficiency within another 2~h~\citep{Hassan2011}. Assuming this time as $t_{\rm cool}$, the practically useful activity will be $A(t_{\rm cool})$$\approx$\textonequarter$A(t_{\rm irr.})$. For our $Y_{\rm meas.}$ at $E_{\alpha}$$=$8~MeV, $t_{\rm irr.}$$=$2~h, and $t_{\rm cool}$$=$2~h, a beam current of  5~mA would result in 210~MBq of $^{68}{\rm Ga}$. Such a He$^{++}$ beam current is within reach of medical isotope-production accelerators under development~\citep[e.g.][]{Nagae2025}. Assuming the natural copper target thickness is 5~$\mu$m, so that $E_{\alpha}$ ranges from 8~MeV down to the $^{65}{\rm Cu}(\alpha,n)$ threshold, the target would sustain a power of 7.5~kW. While this is considerable, it is within the range of standard target-cooling approaches~\citep{iaea2008}. Therefore, radioimpurity-free production of $^{68}{\rm Ga}$ for use as a medical isotope via $^{65}{\rm Cu}(\alpha,n)$ at $E_{\alpha}$$=$8~MeV using natural copper targets is technically feasible, if somewhat impractical.

\section{Cross Section, Deep Underground Detector Backgrounds, and the Astrophysical Reaction Rate}\label{sec:csrate}

Fig.~\ref{fig:CrossSection} shows the $^{65}{\rm Cu}(\alpha,n)$ cross section from all published direct measurements~\citep{Hermanne2015,Levkovskij1991,Bryant1963,Bonesso1991,Scelecsenyi2012,Bhardwaj1988,Singh1994}, compared to the evaluations of \citet{Tarkanyi2019} and \citet{Aslam2021}, the best-fit from \citet{Kudryavtsev2008}, and our Hauser-Feshbach calculations with $\alpha$OMP models 1--8. The experimental data in this figure are all from stacked-foil activation measurements, likely with significant uncertainties and inaccuracies at the lowest $E_{\alpha}$, as discussed. The evaluations of \citet{Tarkanyi2019} and \citet{Aslam2021} rely on these data and thus at low $E_{\alpha}$ also significantly exceed the Hauser-Feshbach estimates that agree with our and previous thick-target yield data from single-foil irradiations (See Sec.~\ref{sec:yield}). \citet{Kudryavtsev2008}, meanwhile, recommend a cross section obtained from a Hauser-Feshbach calculation that best reproduces the cross section data derived from the \citet{Stelson1964} thick-target yield. As our Hauser-Feshbach calculations also agree with this dataset, they are in agreement with the recommended cross section from \citet{Kudryavtsev2008}. The \citet{Kudryavtsev2008} calculation is the basis for the $^{65}{\rm Cu}(\alpha,n)$ contribution to the background for deep underground dark matter and neutrino detectors. An alternative $^{65}{\rm Cu}(\alpha,n)$ background estimate comes from Neutron Calculator Based on Talys (NeuCBOT)~\citep{Westerdale2017}, which also uses Talys calculations to reproduce the measurements of \citet{Stelson1964}. We therefore conclude that this background contribution is accurately accounted for.

The nuclear reaction cross section is a key ingredient for the astrophysical reaction rate,
\begin{equation}
    N_{\rm A}\langle\sigma v\rangle\propto\int\sigma(E)Ee^{-E/k_{\rm B}T}dE,
    \label{eqn:rate}
\end{equation}
where $k_{\rm B}$ is the Boltzmann constant, $T$ is the environment temperature, and $E$ is the reaction energy in the center-of-mass frame. The energy range in which $\sigma(E)$ must be known is where the integrand of Eqn.~\ref{eqn:rate} is largest. The energy bound containing 90\% of the integral is known as the astrophysical window. While the $^{65}{\rm Cu}(\alpha,n)$ reaction rate has been of astrophysical interest for explosive nucleosynthesis in general~\citep{Roughton1983}, modern interest in $(\alpha,n)$ reactions for copper isotopes is $\nu$-driven wind nucleosynthesis of elements from strontium to silver~\citep{Psaltis2022}. Though $^{67}{\rm Cu}(\alpha,n)$ is of particular interest for this site, it is expected that an $\alpha$OMP that accurately captures the $\alpha$$+$$^{65}{\rm Cu}$ interaction should adequately capture the $\alpha$$+$$^{67}{\rm Cu}$ interaction as well. 

\begin{figure}
  \centering
\includegraphics[width=1\columnwidth]{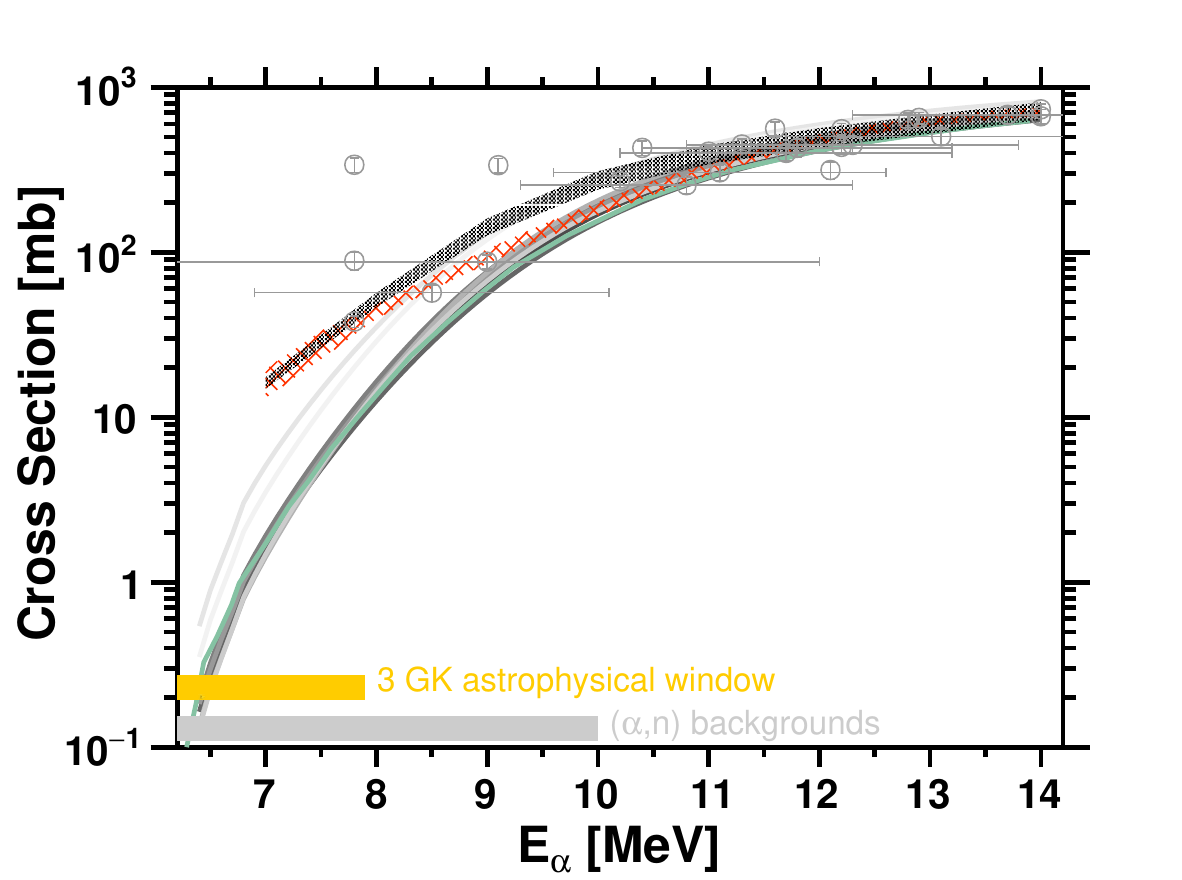}
    \caption{$^{65}{\rm Cu}(\alpha,n)$ cross section calculated with {\tt Talys} $\alpha$OMP models 1--8 (gray lines) compared to recommendations from \citet{Kudryavtsev2008} (green line), \citet{Aslam2021} (black dotted band), \citet{Tarkanyi2019} (red hashed band), and all published direct measurement data~\citep{Hermanne2015,Levkovskij1991,Bryant1963,Bonesso1991,Scelecsenyi2012,Bhardwaj1988,Singh1994}. The orange and gray bands near the bottom of the figure show the energy regions relevant for astrophysics and deep underground detector backgrounds, respectively.}\label{fig:CrossSection}
\end{figure}

\begin{figure}
  \centering
\includegraphics[width=1\columnwidth]{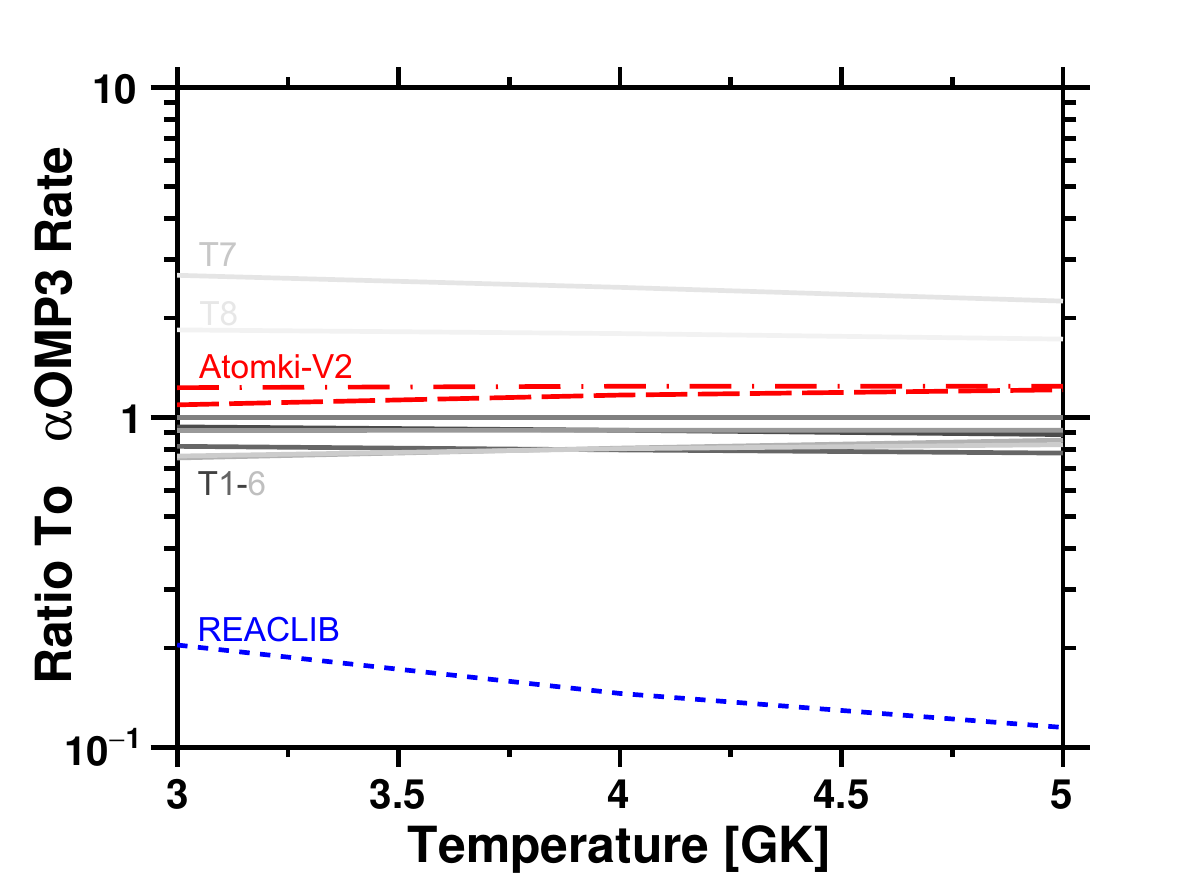}
    \caption{Ratio of astrophysical reaction rates calculated with $\alpha$OMP models 1--8, the Atomki-V2 $\alpha$OMP rates  calculated by \citet{Mohr2021} and \citet{Rauscher2026}, and the recommended rate in the {\tt REACLIB} database~\citep{Cyburt2010}, to the rate calculated with $\alpha$OMP model 3, over the temperature range of interest for $\nu$-driven wind nucleosynthesis.}\label{fig:RateRatio}
\end{figure}

In this spirit, we compare $^{65}$Cu$(\alpha,n)$ astrophysical reaction rates calculated using $\alpha$OMP models 1--8 in {\tt Talys} to other predictions~\citep{Cyburt2010,Mohr2021,Rauscher2026} in Fig.~\ref{fig:RateRatio}. While the {\tt REACLIB} reaction rate library~\citep{Cyburt2010} has often been employed in astrophysics model calculations, it evidently significantly underestimates the $^{65}{\rm Cu}(\alpha,n)$ reaction rate relative to our Hauser-Feshbach calculations. Recently, the Atomki-V2 $\alpha$OMP was developed to improve the description of $\alpha$-induced astrophysical reaction rates~\citep{Mohr2020}. Rates have been calculated with this $\alpha$OMP by \citet{Mohr2021} and \citet{Rauscher2026}, using independent Hauser-Feshbach codes and likely differing in input assumptions; nonetheless, these rates are within 11\% of each other over the 3--5~GK temperature range of interest for $\nu$-driven wind nucleosynthesis. Each are also within a factor of two of the $\alpha$OMP model 1--6 rates recommended here. This is consistent with expectation of \citet{Psaltis2022} that a factor of three variation in the Atomki-V2-derived $(\alpha,n)$ rates is adequate to capture outstanding uncertainties. Based on the thick-target yield comparisons discussed in Sec.~\ref{sec:yield}, we recommend the $^{65}{\rm Cu}(\alpha,n)$ astrophysical reaction rate calculated using {\tt Talys} with $\alpha$OMP model 3, with an uncertainty of 25\% to account for the similar performance of $\alpha$OMP models 1--6. This rate is within 25\% of the Atomki-V2 rates within 3--5~GK, therefore those rates would be equally suitable in $\nu$-driven wind nucleosynthesis calculations.

\section{Conclusions}\label{sec:conclusions}
In conclusion, we performed thick-target yield measurements of $^{65}{\rm Cu}(\alpha,n)^{68}{\rm Ga}$ near threshold, along with Hauser-Feshbach calculations in order to evaluate the impact of our measurements for nuclear medicine, deep-underground measurements of neutrinos and dark matter, and nuclear astrophysics. We determined that the thick-target yield for $^{68}{\rm Ga}$ from $E_{\alpha}$$=$14~MeV $\alpha$-particles impinging on 100\%-enriched copper ($^{65}{\rm Cu}$) is 161$^{+5}_{-24}$~MBq/$\mu$Ah (for He$^{++}$), which is in agreement with the current IAEA estimate, though our yield at $E_{\alpha}$$=$8~MeV is four times lower than the IAEA estimate. We found that the $E_{\alpha}$$=$8~MeV yield is sufficient to enable radioimpurity-free production of $^{68}{\rm Ga}$ for use as a medical isotope via $^{65}{\rm Cu}(\alpha,n)$ using natural copper targets, at least in principle.

Comparisons to our and previous single-foil irradiation data for $^{65}{\rm Cu}(\alpha,n)$ suggest that the $^{65}{\rm Cu}(\alpha,n)$ cross section, and therefore astrophysical reaction rate, is accurately described by Hauser-Feshbach calculations performed using {\tt Talys} $\alpha$OMP models 1--6. These cross sections and rates are in agreement with those currently adopted for deep underground detector backgrounds and for astrophysical nucleosynthesis calculations, respectively. Namely, the cross section recommended by \citet{Kudryavtsev2008} and the rates of \citet{Mohr2021} and \citet{Rauscher2026}. Our results suggest a $^{65}{\rm Cu}(\alpha,n)$ cross section and reaction rate uncertainty of around 25\%, which is likely sufficient for the purposes of detector backgrounds and nucleosynthesis calculations.


\section*{Acknowledgements}
The views expressed are those of the authors and do not reflect the official guidance or position of the United States Government, the United States Department of War, or the United States Air Force. This work was supported in part by the U.S. Department of Energy Office of Science under Grants No. 89243126SSC000179, DE-FG02-88ER40387, and DE-SC0019042 and the U.S. National Nuclear Security Administration through Grants No. DE-NA0003883 and DE-NA0003909. The helium ion source was provided by Grant No. PHY-1827893 from the U.S. National Science Foundation. We also benefited from support by U.S. Department of Energy Office of Science under Grant No. DE-SC0023128 (CeNAM).


\begin{thebibliography}{58}
\expandafter\ifx\csname natexlab\endcsname\relax\def\natexlab#1{#1}\fi
\providecommand{\url}[1]{\texttt{#1}}
\providecommand{\href}[2]{#2}
\providecommand{\path}[1]{#1}
\providecommand{\DOIprefix}{doi:}
\providecommand{\ArXivprefix}{arXiv:}
\providecommand{\URLprefix}{URL: }
\providecommand{\Pubmedprefix}{pmid:}
\providecommand{\doi}[1]{\href{http://dx.doi.org/#1}{\path{#1}}}
\providecommand{\Pubmed}[1]{\href{pmid:#1}{\path{#1}}}
\providecommand{\bibinfo}[2]{#2}
\ifx\xfnm\relax \def\xfnm[#1]{\unskip,\space#1}\fi
\bibitem[{Qaim(2017)}]{Qaim2017}
\bibinfo{author}{S.~M. Qaim},
\newblock \bibinfo{title}{Nuclear data for production and medical application of radionuclides: {P}resent status and future needs},
\newblock \bibinfo{journal}{Nuclear Medicine and Biology} \bibinfo{volume}{44} (\bibinfo{year}{2017}) \bibinfo{pages}{31}. \DOIprefix\doi{10.1016/j.nucmedbio.2016.08.016}.
\bibitem[{T\'{a}rk\'{a}nyi et~al.(2019)T\'{a}rk\'{a}nyi, Ignatyuk, Hermanne, Capote, Carlson, Engle, Kellett, Kib\'{e}di, Kim, Kondev, Hussain, Lebeda, Luca, Nagai, Naik, Nichols, Nortier, Suryanarayana, Tak\'{a}cs, and Verpelli}]{Tarkanyi2019}
\bibinfo{author}{F.~T. T\'{a}rk\'{a}nyi}, \bibinfo{author}{A.~V. Ignatyuk}, \bibinfo{author}{A.~Hermanne}, \bibinfo{author}{R.~Capote}, \bibinfo{author}{B.~V. Carlson}, \bibinfo{author}{J.~W. Engle}, \bibinfo{author}{M.~A. Kellett}, \bibinfo{author}{T.~Kib\'{e}di}, \bibinfo{author}{G.~N. Kim}, \bibinfo{author}{F.~G. Kondev}, \bibinfo{author}{M.~Hussain}, \bibinfo{author}{O.~Lebeda}, \bibinfo{author}{A.~Luca}, \bibinfo{author}{Y.~Nagai}, \bibinfo{author}{H.~Naik}, \bibinfo{author}{A.~L. Nichols}, \bibinfo{author}{F.~M. Nortier}, \bibinfo{author}{S.~V. Suryanarayana}, \bibinfo{author}{S.~Tak\'{a}cs}, \bibinfo{author}{M.~Verpelli},
\newblock \bibinfo{title}{Recommended nuclear data for medical radioisotope production: diagnostic positron emitters},
\newblock \bibinfo{journal}{Journal of Radioanalytical and Nuclear Chemistry} \bibinfo{volume}{319} (\bibinfo{year}{2019}) \bibinfo{pages}{533}. \DOIprefix\doi{10.1007/s10967-018-6380-5}.
\bibitem[{{Kudryavtsev} et~al.(2008){Kudryavtsev}, {Pandola}, and {Tomasello}}]{Kudryavtsev2008}
\bibinfo{author}{V.~A. {Kudryavtsev}}, \bibinfo{author}{L.~{Pandola}}, \bibinfo{author}{V.~{Tomasello}},
\newblock \bibinfo{title}{{Neutron- and muon-induced background in underground physics experiments}},
\newblock \bibinfo{journal}{European Physical Journal A} \bibinfo{volume}{36} (\bibinfo{year}{2008}) \bibinfo{pages}{171--180}. \DOIprefix\doi{10.1140/epja/i2007-10539-6}.
\bibitem[{{Roughton} et~al.(1983){Roughton}, {Intrator}, {Peterson}, {Zaidins}, and {Hansen}}]{Roughton1983}
\bibinfo{author}{N.~A. {Roughton}}, \bibinfo{author}{T.~P. {Intrator}}, \bibinfo{author}{R.~J. {Peterson}}, \bibinfo{author}{C.~S. {Zaidins}}, \bibinfo{author}{C.~J. {Hansen}},
\newblock \bibinfo{title}{{Thick-Target Measurements and Astrophysical Thermonuclear Reaction Rates: Alpha-Induced Reactions}},
\newblock \bibinfo{journal}{Atomic Data and Nuclear Data Tables} \bibinfo{volume}{28} (\bibinfo{year}{1983}) \bibinfo{pages}{341}. \DOIprefix\doi{10.1016/0092-640X(83)90021-9}.
\bibitem[{{Westerdale} and {Meyers}(2017)}]{Westerdale2017}
\bibinfo{author}{S.~{Westerdale}}, \bibinfo{author}{P.~D. {Meyers}},
\newblock \bibinfo{title}{{Radiogenic neutron yield calculations for low-background experiments}},
\newblock \bibinfo{journal}{Nuclear Instruments and Methods in Physics Research A} \bibinfo{volume}{875} (\bibinfo{year}{2017}) \bibinfo{pages}{57--64}. \DOIprefix\doi{10.1016/j.nima.2017.09.007}.
\bibitem[{Baum and Kulkarni(2012)}]{Baum2012}
\bibinfo{author}{R.~P. Baum}, \bibinfo{author}{H.~R. Kulkarni},
\newblock \bibinfo{title}{Theranostics: {F}rom molecular imaging using {G}a-68 labeled tracers and {PET/CT} to personalized radionuclide therapy - {The Bad Berka} experience},
\newblock \bibinfo{journal}{Theranostics} \bibinfo{volume}{2} (\bibinfo{year}{2012}) \bibinfo{pages}{437}. \DOIprefix\doi{10.7150/thno.3645}.
\bibitem[{Banerjee and Pomper(2013)}]{Banerjee2013}
\bibinfo{author}{S.~R. Banerjee}, \bibinfo{author}{M.~G. Pomper},
\newblock \bibinfo{title}{Clinical applications of {G}allium-68},
\newblock \bibinfo{journal}{Applied Radiation and Isotopes} \bibinfo{volume}{76} (\bibinfo{year}{2013}) \bibinfo{pages}{2}. \DOIprefix\doi{10.1016/j.apradiso.2013.01.039}.
\bibitem[{Aslam et~al.(2021)Aslam, Ali, and Hussain}]{Aslam2021}
\bibinfo{author}{M.~T. Aslam}, \bibinfo{author}{W.~Ali}, \bibinfo{author}{M.~Hussain},
\newblock \bibinfo{title}{Nuclear model analysis of the $^{65}${Cu}$(\alpha,n)^{68}${Ga} reaction for the production of $^{68}${Ga} up to 40~{MeV}},
\newblock \bibinfo{journal}{Applied Radiation and Isotopes} \bibinfo{volume}{170} (\bibinfo{year}{2021}) \bibinfo{pages}{109590}. \DOIprefix\doi{10.1016/j.apradiso.2021.109590}.
\bibitem[{{Brandenburg} et~al.(2023){Brandenburg}, {Hamad}, {Meisel}, {Brune}, {Carter}, {deBoer}, {Derkin}, {Feathers}, {Ingram}, {Jones-Alberty}, {Kenady}, {Massey}, {Saxena}, {Soltesz}, {Subedi}, {Voinov}, {Warren}, and {Wiescher}}]{Brandenburg2023a}
\bibinfo{author}{K.~{Brandenburg}}, \bibinfo{author}{G.~{Hamad}}, \bibinfo{author}{Z.~{Meisel}}, \bibinfo{author}{C.~R. {Brune}}, \bibinfo{author}{D.~E. {Carter}}, \bibinfo{author}{R.~J. {deBoer}}, \bibinfo{author}{J.~{Derkin}}, \bibinfo{author}{C.~{Feathers}}, \bibinfo{author}{D.~C. {Ingram}}, \bibinfo{author}{Y.~{Jones-Alberty}}, \bibinfo{author}{B.~{Kenady}}, \bibinfo{author}{T.~N. {Massey}}, \bibinfo{author}{M.~{Saxena}}, \bibinfo{author}{D.~{Soltesz}}, \bibinfo{author}{S.~K. {Subedi}}, \bibinfo{author}{A.~V. {Voinov}}, \bibinfo{author}{J.~{Warren}}, \bibinfo{author}{M.~{Wiescher}},
\newblock \bibinfo{title}{{Measurements of the $^{13}$C({\ensuremath{\alpha}} ,n )$^{16}$O cross section up to E$_{{\ensuremath{\alpha}}}$=8 MeV}},
\newblock \bibinfo{journal}{Physical Review C Letters} \bibinfo{volume}{108} (\bibinfo{year}{2023}) \bibinfo{pages}{L061601}. \DOIprefix\doi{10.1103/PhysRevC.108.L061601}.
\bibitem[{{Rauscher}(2010)}]{Rauscher2010}
\bibinfo{author}{T.~{Rauscher}},
\newblock \bibinfo{title}{{Relevant energy ranges for astrophysical reaction rates}},
\newblock \bibinfo{journal}{Physical Review C} \bibinfo{volume}{81} (\bibinfo{year}{2010}) \bibinfo{pages}{045807}. \DOIprefix\doi{10.1103/PhysRevC.81.045807}.
\bibitem[{Szelecs\'{e}nyi et~al.(2012)Szelecs\'{e}nyi, Kov\'{a}cs, Nagatsu, Fukumura, Suzuki, and Mukai}]{Scelecsenyi2012}
\bibinfo{author}{F.~Szelecs\'{e}nyi}, \bibinfo{author}{Z.~Kov\'{a}cs}, \bibinfo{author}{K.~Nagatsu}, \bibinfo{author}{K.~Fukumura}, \bibinfo{author}{K.~Suzuki}, \bibinfo{author}{K.~Mukai},
\newblock \bibinfo{title}{Investigation of direct production of $^{68}${G}a with low energy multiparticle accelerator},
\newblock \bibinfo{journal}{Radiochimica Acta} \bibinfo{volume}{100} (\bibinfo{year}{2012}) \bibinfo{pages}{5}. \DOIprefix\doi{10.1524/ract.2011.1896}.
\bibitem[{{Wang} et~al.(2021){Wang}, {Huang}, {Kondev}, {Audi}, and {Naimi}}]{Wang2021}
\bibinfo{author}{M.~{Wang}}, \bibinfo{author}{W.~J. {Huang}}, \bibinfo{author}{F.~G. {Kondev}}, \bibinfo{author}{G.~{Audi}}, \bibinfo{author}{S.~{Naimi}},
\newblock \bibinfo{title}{{The AME 2020 atomic mass evaluation (II). Tables, graphs and references}},
\newblock \bibinfo{journal}{Chinese Physics C} \bibinfo{volume}{45} (\bibinfo{year}{2021}) \bibinfo{pages}{030003}. \DOIprefix\doi{10.1088/1674-1137/abddaf}.
\bibitem[{{Bryant} et~al.(1963){Bryant}, {Cochran}, and {Knight}}]{Bryant1963}
\bibinfo{author}{E.~A. {Bryant}}, \bibinfo{author}{D.~R. {Cochran}}, \bibinfo{author}{J.~D. {Knight}},
\newblock \bibinfo{title}{{Excitation Functions of Reactions of 7- to 24-MeV He$^{3}$ Ions with Cu$^{63}$ and Cu$^{65}$}},
\newblock \bibinfo{journal}{Physical Review} \bibinfo{volume}{130} (\bibinfo{year}{1963}) \bibinfo{pages}{1512--1522}. \DOIprefix\doi{10.1103/PhysRev.130.1512}.
\bibitem[{{Stelson} and {McGowan}(1964)}]{Stelson1964}
\bibinfo{author}{P.~H. {Stelson}}, \bibinfo{author}{F.~K. {McGowan}},
\newblock \bibinfo{title}{{Cross Sections for ({\ensuremath{\alpha}},n) Reactions for Medium-Weight Nuclei}},
\newblock \bibinfo{journal}{Physical Review} \bibinfo{volume}{133} (\bibinfo{year}{1964}) \bibinfo{pages}{911--919}. \DOIprefix\doi{10.1103/PhysRev.133.B911}.
\bibitem[{Dmitriev(1986)}]{Dmitriev1986}
\bibinfo{author}{P.~P. Dmitriev}, \bibinfo{title}{{{\it IAEA translation of} Radionuclide Yield in Reactions with Protons, Deuterons, Alpha Particles and Helium-3 (Handbook), {INDC(CCP)-263/G+CN+SZ}}}, \bibinfo{publisher}{International Atomic Energy Agency}, \bibinfo{address}{Vienna, Austria}, \bibinfo{year}{1986}. \URLprefix \url{https://nds.iaea.org/records/9dpgh-r8m04}.
\bibitem[{{Bhardwaj} et~al.(1988){Bhardwaj}, {Gautam}, and {Prasad}}]{Bhardwaj1988}
\bibinfo{author}{H.~D. {Bhardwaj}}, \bibinfo{author}{A.~K. {Gautam}}, \bibinfo{author}{R.~{Prasad}},
\newblock \bibinfo{title}{{Measurement and analysis of excitation functions for alpha-induced reactions in copper}},
\newblock \bibinfo{journal}{Pramana} \bibinfo{volume}{31} (\bibinfo{year}{1988}) \bibinfo{pages}{109--123}. \DOIprefix\doi{10.1007/BF02846965}.
\bibitem[{Levkovskij(1991)}]{Levkovskij1991}
\bibinfo{author}{V.~N. Levkovskij}, \bibinfo{title}{{Middle Mass Nuclides (A = 40-100) Activation Cross Sections By Medium Energy (E= 10-50 MeV) Protons and Alpha Particles (Experiment and Systematics)}}, \bibinfo{publisher}{Inter-vesi}, \bibinfo{address}{Moscow, Russia}, \bibinfo{year}{1991}.
\bibitem[{{Bonesso} et~al.(1991){Bonesso}, Ozafran, Mosca, Vazquez, Capurro, and Nassiff}]{Bonesso1991}
\bibinfo{author}{O.~{Bonesso}}, \bibinfo{author}{M.~J. Ozafran}, \bibinfo{author}{H.~O. Mosca}, \bibinfo{author}{M.~E. Vazquez}, \bibinfo{author}{O.~A. Capurro}, \bibinfo{author}{S.~J. Nassiff},
\newblock \bibinfo{title}{Study of pre-equilibrium effects on $\alpha$-induced reactions on copper},
\newblock \bibinfo{journal}{Journal of Radioanalytical and Nuclear Chemistry} \bibinfo{volume}{152} (\bibinfo{year}{1991}) \bibinfo{pages}{189}. \DOIprefix\doi{10.1007/BF02042152}.
\bibitem[{{Singh} et~al.(1994){Singh}, {Patel}, {Somayajulu}, and {Chintalapudi}}]{Singh1994}
\bibinfo{author}{N.~L. {Singh}}, \bibinfo{author}{B.~J. {Patel}}, \bibinfo{author}{D.~R.~S. {Somayajulu}}, \bibinfo{author}{S.~N. {Chintalapudi}},
\newblock \bibinfo{title}{{Analysis of the excitation functions of ({\ensuremath{\alpha}}, xnyp) reactions on natural copper}},
\newblock \bibinfo{journal}{Pramana} \bibinfo{volume}{42} (\bibinfo{year}{1994}) \bibinfo{pages}{349--363}. \DOIprefix\doi{10.1007/BF02847761}.
\bibitem[{{Hermanne} et~al.(2015){Hermanne}, {Adam Rebeles}, {T{\'a}rk{\'a}nyi}, and {Tak{\'a}cs}}]{Hermanne2015}
\bibinfo{author}{A.~{Hermanne}}, \bibinfo{author}{R.~{Adam Rebeles}}, \bibinfo{author}{F.~{T{\'a}rk{\'a}nyi}}, \bibinfo{author}{S.~{Tak{\'a}cs}},
\newblock \bibinfo{title}{{Alpha particle induced reactions on $^{nat}$Cr up to 39 MeV: Experimental cross-sections, comparison with theoretical calculations and thick target yields for medically relevant $^{52g}$Fe production}},
\newblock \bibinfo{journal}{Nuclear Instruments and Methods in Physics Research B} \bibinfo{volume}{356} (\bibinfo{year}{2015}) \bibinfo{pages}{28--41}. \DOIprefix\doi{10.1016/j.nimb.2015.04.025}.
\bibitem[{{Paul} and {Schinner}(2005)}]{Paul2005}
\bibinfo{author}{H.~{Paul}}, \bibinfo{author}{A.~{Schinner}},
\newblock \bibinfo{title}{{Judging the reliability of stopping power tables and programs for protons and alpha particles using statistical methods}},
\newblock \bibinfo{journal}{Nuclear Instruments and Methods in Physics Research B} \bibinfo{volume}{227} (\bibinfo{year}{2005}) \bibinfo{pages}{461--470}. \DOIprefix\doi{10.1016/j.nimb.2004.10.007}.
\bibitem[{{Montanari} and {Dimitriou}(2017)}]{Montanari2017}
\bibinfo{author}{C.~C. {Montanari}}, \bibinfo{author}{P.~{Dimitriou}},
\newblock \bibinfo{title}{{The IAEA stopping power database, following the trends in stopping power of ions in matter}},
\newblock \bibinfo{journal}{Nuclear Instruments and Methods in Physics Research B} \bibinfo{volume}{408} (\bibinfo{year}{2017}) \bibinfo{pages}{50--55}. \DOIprefix\doi{10.1016/j.nimb.2017.03.138}.
\bibitem[{{Hauser} and {Feshbach}(1952)}]{Hauser1952}
\bibinfo{author}{W.~{Hauser}}, \bibinfo{author}{H.~{Feshbach}},
\newblock \bibinfo{title}{{The Inelastic Scattering of Neutrons}},
\newblock \bibinfo{journal}{Physical Review} \bibinfo{volume}{87} (\bibinfo{year}{1952}) \bibinfo{pages}{366--373}. \DOIprefix\doi{10.1103/PhysRev.87.366}.
\bibitem[{Kamal et~al.(2024)Kamal, Nizam, and Abdul~Aziz}]{Kamal2024}
\bibinfo{author}{N.~Kamal}, \bibinfo{author}{S.~Nizam}, \bibinfo{author}{A.~Abdul~Aziz},
\newblock \bibinfo{title}{The effects of nuclear level density model and alpha optical model potential to the excitation functions of novel therapeutic radionuclides},
\newblock \bibinfo{journal}{Applied Radiation and Isotopes} \bibinfo{volume}{203} (\bibinfo{year}{2024}) \bibinfo{pages}{111085}. \DOIprefix\doi{10.1016/j.apradiso.2023.111085}.
\bibitem[{Meisel et~al.(2017)Meisel, Brune, Grimes, Ingram, Massey, and Voinov}]{Meisel2017}
\bibinfo{author}{Z.~Meisel}, \bibinfo{author}{C.~Brune}, \bibinfo{author}{S.~Grimes}, \bibinfo{author}{D.~Ingram}, \bibinfo{author}{T.~Massey}, \bibinfo{author}{A.~Voinov},
\newblock \bibinfo{title}{{The Edwards Accelerator Laboratory at Ohio University}},
\newblock \bibinfo{journal}{Physics Procedia} \bibinfo{volume}{90} (\bibinfo{year}{2017}) \bibinfo{pages}{448--454}. \DOIprefix\doi{10.1016/j.phpro.2017.09.050}.
\bibitem[{{Brandenburg} et~al.(2022){Brandenburg}, {Hamad}, {Meisel}, {Brune}, {Carter}, {Danley}, {Derkin}, {Jones-Alberty}, {Kenady}, {Massey}, {Paneru}, {Saxena}, {Soltesz}, {Subedi}, and {Warren}}]{Brandenburg2022}
\bibinfo{author}{K.~{Brandenburg}}, \bibinfo{author}{G.~{Hamad}}, \bibinfo{author}{Z.~{Meisel}}, \bibinfo{author}{C.~R. {Brune}}, \bibinfo{author}{D.~E. {Carter}}, \bibinfo{author}{T.~{Danley}}, \bibinfo{author}{J.~{Derkin}}, \bibinfo{author}{Y.~{Jones-Alberty}}, \bibinfo{author}{B.~{Kenady}}, \bibinfo{author}{T.~N. {Massey}}, \bibinfo{author}{S.~{Paneru}}, \bibinfo{author}{M.~{Saxena}}, \bibinfo{author}{D.~{Soltesz}}, \bibinfo{author}{S.~K. {Subedi}}, \bibinfo{author}{J.~{Warren}},
\newblock \bibinfo{title}{{The $^{3}$He BF$_{3}$ Giant Barrel (HeBGB) neutron detector}},
\newblock \bibinfo{journal}{Journal of Instrumentation} \bibinfo{volume}{17} (\bibinfo{year}{2022}) \bibinfo{pages}{P05004}. \DOIprefix\doi{10.1088/1748-0221/17/05/P05004}.
\bibitem[{{Brandenburg} et~al.(2023){Brandenburg}, {Hamad}, {Meisel}, {Brune}, {Carter}, {Derkin}, {Ingram}, {Jones-Alberty}, {Kenady}, {Massey}, {Saxena}, {Soltesz}, {Subedi}, and {Warren}}]{Brandenburg2023b}
\bibinfo{author}{K.~{Brandenburg}}, \bibinfo{author}{G.~{Hamad}}, \bibinfo{author}{Z.~{Meisel}}, \bibinfo{author}{C.~R. {Brune}}, \bibinfo{author}{D.~E. {Carter}}, \bibinfo{author}{J.~{Derkin}}, \bibinfo{author}{D.~C. {Ingram}}, \bibinfo{author}{Y.~{Jones-Alberty}}, \bibinfo{author}{B.~{Kenady}}, \bibinfo{author}{T.~N. {Massey}}, \bibinfo{author}{M.~{Saxena}}, \bibinfo{author}{D.~{Soltesz}}, \bibinfo{author}{S.~K. {Subedi}}, \bibinfo{author}{J.~{Warren}},
\newblock \bibinfo{title}{{Measurements of the $^{27}$Al({\ensuremath{\alpha}},n) Thick Target Yield near Threshold}},
\newblock \bibinfo{journal}{Nuclear Science and Engineering} \bibinfo{volume}{197} (\bibinfo{year}{2023}) \bibinfo{pages}{510--516}. \DOIprefix\doi{10.1080/00295639.2022.2118483}.
\bibitem[{{Douglas} and {MacDonald}(1959)}]{Douglas1959}
\bibinfo{author}{A.~C. {Douglas}}, \bibinfo{author}{N.~{MacDonald}},
\newblock \bibinfo{title}{{Compound nucleus processes in medium mass nuclei}},
\newblock \bibinfo{journal}{Nuclear Physics} \bibinfo{volume}{13} (\bibinfo{year}{1959}) \bibinfo{pages}{382--396}. \DOIprefix\doi{10.1016/0029-5582(59)90116-6}.
\bibitem[{{Grimes} et~al.(1974){Grimes}, {Anderson}, {McClure}, {Pohl}, and {Wong}}]{Grimes1974}
\bibinfo{author}{S.~M. {Grimes}}, \bibinfo{author}{J.~D. {Anderson}}, \bibinfo{author}{J.~W. {McClure}}, \bibinfo{author}{B.~A. {Pohl}}, \bibinfo{author}{C.~{Wong}},
\newblock \bibinfo{title}{{Level density and spin cutoff parameters from continuum (p,n) and ({\ensuremath{\alpha}},n) spectra}},
\newblock \bibinfo{journal}{Physical Review C} \bibinfo{volume}{10} (\bibinfo{year}{1974}) \bibinfo{pages}{2373--2386}. \DOIprefix\doi{10.1103/PhysRevC.10.2373}.
\bibitem[{{Blondiaux} et~al.(1984){Blondiaux}, {Valladon}, {Quaglia}, {Robaye}, {Weber}, and {Debrun}}]{Blondiaux1984}
\bibinfo{author}{G.~{Blondiaux}}, \bibinfo{author}{M.~{Valladon}}, \bibinfo{author}{L.~{Quaglia}}, \bibinfo{author}{G.~{Robaye}}, \bibinfo{author}{G.~{Weber}}, \bibinfo{author}{J.~L. {Debrun}},
\newblock \bibinfo{title}{{Study of the growth of carbon on targets during ion bombardment}},
\newblock \bibinfo{journal}{Nuclear Instruments and Methods in Physics Research A} \bibinfo{volume}{227} (\bibinfo{year}{1984}) \bibinfo{pages}{19--23}. \DOIprefix\doi{10.1016/0168-9002(84)90095-0}.
\bibitem[{Berglund and Wieser(2011)}]{Berglund2011}
\bibinfo{author}{M.~Berglund}, \bibinfo{author}{M.~E. Wieser},
\newblock \bibinfo{title}{Isotopic compositions of the elements 2009 {(IUPAC Technical Report)}},
\newblock \bibinfo{journal}{Pure and Applied Chemistry} \bibinfo{volume}{83} (\bibinfo{year}{2011}) \bibinfo{pages}{397}. \DOIprefix\doi{10.1351/PAC-REP-10-06-02}.
\bibitem[{{Koning} et~al.(2023){Koning}, {Hilaire}, and {Goriely}}]{Koning2023}
\bibinfo{author}{A.~{Koning}}, \bibinfo{author}{S.~{Hilaire}}, \bibinfo{author}{S.~{Goriely}},
\newblock \bibinfo{title}{{TALYS: modeling of nuclear reactions}},
\newblock \bibinfo{journal}{European Physical Journal A} \bibinfo{volume}{59} (\bibinfo{year}{2023}) \bibinfo{pages}{131}. \DOIprefix\doi{10.1140/epja/s10050-023-01034-3}.
\bibitem[{{Ziegler} et~al.(2010){Ziegler}, {Ziegler}, and {Biersack}}]{Ziegler2010}
\bibinfo{author}{J.~F. {Ziegler}}, \bibinfo{author}{M.~D. {Ziegler}}, \bibinfo{author}{J.~P. {Biersack}},
\newblock \bibinfo{title}{{SRIM - The stopping and range of ions in matter (2010)}},
\newblock \bibinfo{journal}{Nuclear Instruments and Methods in Physics Research B} \bibinfo{volume}{268} (\bibinfo{year}{2010}) \bibinfo{pages}{1818--1823}. \DOIprefix\doi{10.1016/j.nimb.2010.02.091}.
\bibitem[{{Mohr}(2015)}]{Mohr2015}
\bibinfo{author}{P.~{Mohr}},
\newblock \bibinfo{title}{{Cross sections of {\ensuremath{\alpha}}-induced reactions for targets with masses A {\ensuremath{\approx}} 20-50 at low energies}},
\newblock \bibinfo{journal}{European Physical Journal A} \bibinfo{volume}{51} (\bibinfo{year}{2015}) \bibinfo{pages}{56}. \DOIprefix\doi{10.1140/epja/i2015-15056-5}.
\bibitem[{{Rauscher}(2012)}]{Rauscher2012}
\bibinfo{author}{T.~{Rauscher}},
\newblock \bibinfo{title}{{Sensitivity of Astrophysical Reaction Rates to Nuclear Uncertainties}},
\newblock \bibinfo{journal}{Astrophysical Journal Supplement Series} \bibinfo{volume}{201} (\bibinfo{year}{2012}) \bibinfo{pages}{26}. \DOIprefix\doi{10.1088/0067-0049/201/2/26}.
\bibitem[{{McFadden} and {Satchler}(1966)}]{McFadden1966}
\bibinfo{author}{L.~{McFadden}}, \bibinfo{author}{G.~R. {Satchler}},
\newblock \bibinfo{title}{{Optical-model analysis of the scattering of 24.7 MeV alpha particles}},
\newblock \bibinfo{journal}{Nuclear Physics} \bibinfo{volume}{84} (\bibinfo{year}{1966}) \bibinfo{pages}{177--200}. \DOIprefix\doi{10.1016/0029-5582(66)90441-X}.
\bibitem[{{Demetriou} et~al.(2002){Demetriou}, {Grama}, and {Goriely}}]{Demetriou2002}
\bibinfo{author}{P.~{Demetriou}}, \bibinfo{author}{C.~{Grama}}, \bibinfo{author}{S.~{Goriely}},
\newblock \bibinfo{title}{{Improved global {\ensuremath{\alpha}}-optical model potentials at low energies}},
\newblock \bibinfo{journal}{Nuclear Physics A} \bibinfo{volume}{707} (\bibinfo{year}{2002}) \bibinfo{pages}{253--276}. \DOIprefix\doi{10.1016/S0375-9474(02)00756-X}.
\bibitem[{{Avrigeanu} et~al.(2014){Avrigeanu}, {Avrigeanu}, and {M{\v{a}}n{\v{a}}ilescu}}]{Avrigeanu2014}
\bibinfo{author}{V.~{Avrigeanu}}, \bibinfo{author}{M.~{Avrigeanu}}, \bibinfo{author}{C.~{M{\v{a}}n{\v{a}}ilescu}},
\newblock \bibinfo{title}{{Further explorations of the {\ensuremath{\alpha}}-particle optical model potential at low energies for the mass range A{\ensuremath{\approx}}45-209}},
\newblock \bibinfo{journal}{Physical Review C} \bibinfo{volume}{90} (\bibinfo{year}{2014}) \bibinfo{pages}{044612}. \DOIprefix\doi{10.1103/PhysRevC.90.044612}.
\bibitem[{{Nolte} et~al.(1987){Nolte}, {Machner}, and {Bojowald}}]{Nolte1987}
\bibinfo{author}{M.~{Nolte}}, \bibinfo{author}{H.~{Machner}}, \bibinfo{author}{J.~{Bojowald}},
\newblock \bibinfo{title}{{Global optical potential for {\ensuremath{\alpha}} particles with energies above 80 MeV}},
\newblock \bibinfo{journal}{Physical Review C} \bibinfo{volume}{36} (\bibinfo{year}{1987}) \bibinfo{pages}{1312--1316}. \DOIprefix\doi{10.1103/PhysRevC.36.1312}.
\bibitem[{{Avrigeanu} et~al.(1994){Avrigeanu}, {Hodgson}, and {Avrigeanu}}]{Avrigeanu1994}
\bibinfo{author}{V.~{Avrigeanu}}, \bibinfo{author}{P.~E. {Hodgson}}, \bibinfo{author}{M.~{Avrigeanu}},
\newblock \bibinfo{title}{{Global optical potentials for emitted alpha particles}},
\newblock \bibinfo{journal}{Physical Review C} \bibinfo{volume}{49} (\bibinfo{year}{1994}) \bibinfo{pages}{2136--2141}. \DOIprefix\doi{10.1103/PhysRevC.49.2136}.
\bibitem[{{Andersen} et~al.(1977){Andersen}, {Bak}, {Knudsen}, and {Nielsen}}]{Andersen1977}
\bibinfo{author}{H.~H. {Andersen}}, \bibinfo{author}{J.~F. {Bak}}, \bibinfo{author}{H.~{Knudsen}}, \bibinfo{author}{B.~R. {Nielsen}},
\newblock \bibinfo{title}{{Stopping power of Al, Cu, Ag, and Au for MeV hydrogen, helium, and lithium ions. Z$^{3}$$_{1}$ and Z$^{4}$$_{1}$ proportional deviations from the Bethe formula}},
\newblock \bibinfo{journal}{Physical Review A} \bibinfo{volume}{16} (\bibinfo{year}{1977}) \bibinfo{pages}{1929--1940}. \DOIprefix\doi{10.1103/PhysRevA.16.1929}.
\bibitem[{{Ishiwari} et~al.(1975){Ishiwari}, {Shiomi}, {Katayama-Kinoshita}, and {Sawada-Yasue}}]{Ishiwari1975}
\bibinfo{author}{R.~{Ishiwari}}, \bibinfo{author}{N.~{Shiomi}}, \bibinfo{author}{T.~{Katayama-Kinoshita}}, \bibinfo{author}{F.~{Sawada-Yasue}},
\newblock \bibinfo{title}{{Search for Possible Geometrical Effect on Stopping Power Measurement}},
\newblock \bibinfo{journal}{Journal of the Physical Society of Japan} \bibinfo{volume}{39} (\bibinfo{year}{1975}) \bibinfo{pages}{557--565}. \DOIprefix\doi{10.1143/JPSJ.39.557}.
\bibitem[{{Ishiwari} et~al.(1977){Ishiwari}, {Shiomi}, {Kinoshita}, and {Yasue}}]{Ishiwari1977}
\bibinfo{author}{R.~{Ishiwari}}, \bibinfo{author}{N.~{Shiomi}}, \bibinfo{author}{T.~{Kinoshita}}, \bibinfo{author}{F.~{Yasue}},
\newblock \bibinfo{title}{Stopping powers of {Al}, {Ni}, {Cu}, {Ag}, and {Ta} for 8.78 {MeV} alpha particles},
\newblock \bibinfo{journal}{Bulletin of the Institute for Chemical Research, Kyoto University} \bibinfo{volume}{55} (\bibinfo{year}{1977}) \bibinfo{pages}{68}. \URLprefix \url{http://hdl.handle.net/2433/75168}.
\bibitem[{Berger et~al.(2017)Berger, Coursey, Zucker, and Chang}]{Berger2017}
\bibinfo{author}{M.~J. Berger}, \bibinfo{author}{J.~S. Coursey}, \bibinfo{author}{M.~A. Zucker}, \bibinfo{author}{J.~Chang}, \bibinfo{title}{Stopping-Power \& Range Tables for Electrons, Protons, and Helium Ions}, \bibinfo{type}{Technical Report}, National Institute of Standards and Technologies, \bibinfo{year}{2017}. \DOIprefix\doi{10.18434/T4NC7P}.
\bibitem[{{Bivort Haiek} et~al.(2022){Bivort Haiek}, {Mendez}, {Montanari}, and {Mitnik}}]{Bivort2022}
\bibinfo{author}{F.~{Bivort Haiek}}, \bibinfo{author}{A.~M.~P. {Mendez}}, \bibinfo{author}{C.~C. {Montanari}}, \bibinfo{author}{D.~M. {Mitnik}},
\newblock \bibinfo{title}{{ESPNN: A novel electronic stopping power neural-network code built on the IAEA stopping power database. I. Atomic targets}},
\newblock \bibinfo{journal}{Journal of Applied Physics} \bibinfo{volume}{132} (\bibinfo{year}{2022}) \bibinfo{pages}{245103}. \DOIprefix\doi{10.1063/5.0130875}.
\bibitem[{Leo(1994)}]{Leo1994}
\bibinfo{author}{W.~R. Leo}, \bibinfo{title}{Techniques for Nuclear and Particle Physics Experiments}, \bibinfo{publisher}{Springer-Verlag Berlin Heidelberg}, \bibinfo{address}{New York, USA}, \bibinfo{year}{1994}.
\bibitem[{{Mohr} et~al.(2020){Mohr}, {F{\"u}l{\"o}p}, {Gy{\"u}rky}, {Kiss}, and {Sz{\"u}cs}}]{Mohr2020}
\bibinfo{author}{P.~{Mohr}}, \bibinfo{author}{Z.~{F{\"u}l{\"o}p}}, \bibinfo{author}{G.~{Gy{\"u}rky}}, \bibinfo{author}{G.~G. {Kiss}}, \bibinfo{author}{T.~{Sz{\"u}cs}},
\newblock \bibinfo{title}{{Successful Prediction of Total {\ensuremath{\alpha}} -Induced Reaction Cross Sections at Astrophysically Relevant Sub-Coulomb Energies Using a Novel Approach}},
\newblock \bibinfo{journal}{Physical Review Letters} \bibinfo{volume}{124} (\bibinfo{year}{2020}) \bibinfo{pages}{252701}. \DOIprefix\doi{10.1103/PhysRevLett.124.252701}.
\bibitem[{{International Atomic Energy Agency}(2001)}]{iaea2001}
\bibinfo{author}{{International Atomic Energy Agency}}, \bibinfo{title}{Charged Particle Cross-Section Database for Medical Radioisotope Production: Diagnostic Radioisotopes and Monitor Reactions}, number \bibinfo{number}{1211} in \bibinfo{series}{IAEA-TECDOC}, \bibinfo{publisher}{{International Atomic Energy Agency}}, \bibinfo{address}{Vienna}, \bibinfo{year}{2001}. \URLprefix \url{http://www.iaea.org/publications/6050/charged-particle-cross-section-database-for-medical-radioisotope-production-diagnostic-radioisotopes-and-monitor-reactions}.
\bibitem[{Otuka and Tak\'{a}cs(2015)}]{Otuka2015}
\bibinfo{author}{N.~Otuka}, \bibinfo{author}{S.~Tak\'{a}cs},
\newblock \bibinfo{title}{Definitions of radioisotope thick target yields},
\newblock \bibinfo{journal}{Radiochimica Acta} \bibinfo{volume}{103} (\bibinfo{year}{2015}) \bibinfo{pages}{1--6}. \URLprefix \url{https://doi.org/10.1515/ract-2013-2234}. \DOIprefix\doi{doi:10.1515/ract-2013-2234}.
\bibitem[{Walker et~al.(2013)Walker, Smith, Liu, Moore, Clanton, and Stabin}]{Walker2013}
\bibinfo{author}{R.~C. Walker}, \bibinfo{author}{G.~T. Smith}, \bibinfo{author}{E.~Liu}, \bibinfo{author}{B.~Moore}, \bibinfo{author}{J.~Clanton}, \bibinfo{author}{M.~Stabin},
\newblock \bibinfo{title}{Measured human dosimetry of {68Ga-DOTATATE}},
\newblock \bibinfo{journal}{Journal of Nuclear Medicine} \bibinfo{volume}{54} (\bibinfo{year}{2013}) \bibinfo{pages}{855--860}. \URLprefix \url{https://jnm.snmjournals.org/content/54/6/855}. \DOIprefix\doi{10.2967/jnumed.112.114165}.
\bibitem[{{McCutchan}(2012)}]{Mccutchan2012}
\bibinfo{author}{E.~A. {McCutchan}},
\newblock \bibinfo{title}{{Nuclear Data Sheets for A = 68}},
\newblock \bibinfo{journal}{Nuclear Data Sheets} \bibinfo{volume}{113} (\bibinfo{year}{2012}) \bibinfo{pages}{1735--1870}. \DOIprefix\doi{10.1016/j.nds.2012.06.002}.
\bibitem[{Hassan(2011)}]{Hassan2011}
\bibinfo{author}{K.~F. Hassan},
\newblock \bibinfo{title}{Rapid separation of radiogallium from {Zn} and {Cu} targets using anion exchange technique},
\newblock \bibinfo{journal}{Journal of Radioanalytical and Nuclear Chemistry} \bibinfo{volume}{289} (\bibinfo{year}{2011}) \bibinfo{pages}{801}. \DOIprefix\doi{10.1007/s10967-011-1145-4}.
\bibitem[{Nagae et~al.(2025)Nagae, Murata, Ikeda, Kikuchi, Yoshimura, Ma, Narita, and Hayashizaki}]{Nagae2025}
\bibinfo{author}{D.~Nagae}, \bibinfo{author}{A.~Murata}, \bibinfo{author}{S.~Ikeda}, \bibinfo{author}{S.~Kikuchi}, \bibinfo{author}{R.~Yoshimura}, \bibinfo{author}{Y.~Ma}, \bibinfo{author}{D.~Narita}, \bibinfo{author}{N.~Hayashizaki},
\newblock \bibinfo{title}{Conceptual design of a {He} beam accelerator system for $^{211}${At} production},
\newblock \bibinfo{journal}{Progress in Nuclear Science and Technology} \bibinfo{volume}{7} (\bibinfo{year}{2025}) \bibinfo{pages}{357}. \DOIprefix\doi{10.15669/pnst.7.357}.
\bibitem[{{International Atomic Energy Agency}(2008)}]{iaea2008}
\bibinfo{author}{{International Atomic Energy Agency}}, \bibinfo{title}{Cyclotron Produced Radionuclides: Principles and Practice}, number \bibinfo{number}{465} in \bibinfo{series}{Technical Reports Series}, \bibinfo{publisher}{{International Atomic Energy Agency}}, \bibinfo{address}{Vienna}, \bibinfo{year}{2008}. \URLprefix \url{http://www.iaea.org/publications/7849/cyclotron-produced-radionuclides-principles-and-practice}.
\bibitem[{{Psaltis} et~al.(2022){Psaltis}, {Arcones}, {Montes}, {Mohr}, {Hansen}, {Jacobi}, and {Schatz}}]{Psaltis2022}
\bibinfo{author}{A.~{Psaltis}}, \bibinfo{author}{A.~{Arcones}}, \bibinfo{author}{F.~{Montes}}, \bibinfo{author}{P.~{Mohr}}, \bibinfo{author}{C.~J. {Hansen}}, \bibinfo{author}{M.~{Jacobi}}, \bibinfo{author}{H.~{Schatz}},
\newblock \bibinfo{title}{{Constraining Nucleosynthesis in Neutrino-driven Winds: Observations, Simulations, and Nuclear Physics}},
\newblock \bibinfo{journal}{Astrophysical Journal} \bibinfo{volume}{935} (\bibinfo{year}{2022}) \bibinfo{pages}{27}. \DOIprefix\doi{10.3847/1538-4357/ac7da7}.
\bibitem[{Mohr et~al.(2021)Mohr, F\"{u}l\"{o}p, Gy\"{u}rky, Kiss, Sz\"{u}cs, Arcones, Jacobi, and Psaltis}]{Mohr2021}
\bibinfo{author}{P.~Mohr}, \bibinfo{author}{Z.~F\"{u}l\"{o}p}, \bibinfo{author}{G.~Gy\"{u}rky}, \bibinfo{author}{G.~G. Kiss}, \bibinfo{author}{T.~Sz\"{u}cs}, \bibinfo{author}{A.~Arcones}, \bibinfo{author}{M.~Jacobi}, \bibinfo{author}{A.~Psaltis},
\newblock \bibinfo{title}{Astrophysical reaction rates of $\alpha$-induced reaction rates for nuclei with 26$\leq${Z}$\leq$83 from the new {A}tomki-{V}2 $\alpha$-nucleus potential},
\newblock \bibinfo{journal}{Atomic Data and Nuclear Data Tables} \bibinfo{volume}{142} (\bibinfo{year}{2021}) \bibinfo{pages}{101453}. \DOIprefix\doi{10.1016/j.adt.2021.101453}.
\bibitem[{{Rauscher}(2026)}]{Rauscher2026}
\bibinfo{author}{T.~{Rauscher}},
\newblock \bibinfo{title}{{Astrophysical reaction rates for charged-particle induced reactions on proton-rich nuclides}},
\newblock \bibinfo{journal}{European Physical Journal A} \bibinfo{volume}{62} (\bibinfo{year}{2026}) \bibinfo{pages}{35}. \DOIprefix\doi{10.1140/epja/s10050-026-01809-4}.
\bibitem[{{Cyburt} et~al.(2010){Cyburt}, {Amthor}, {Ferguson}, {Meisel}, {Smith}, {Warren}, {Heger}, {Hoffman}, {Rauscher}, {Sakharuk}, {Schatz}, {Thielemann}, and {Wiescher}}]{Cyburt2010}
\bibinfo{author}{R.~H. {Cyburt}}, \bibinfo{author}{A.~M. {Amthor}}, \bibinfo{author}{R.~{Ferguson}}, \bibinfo{author}{Z.~{Meisel}}, \bibinfo{author}{K.~{Smith}}, \bibinfo{author}{S.~{Warren}}, \bibinfo{author}{A.~{Heger}}, \bibinfo{author}{R.~D. {Hoffman}}, \bibinfo{author}{T.~{Rauscher}}, \bibinfo{author}{A.~{Sakharuk}}, \bibinfo{author}{H.~{Schatz}}, \bibinfo{author}{F.~K. {Thielemann}}, \bibinfo{author}{M.~{Wiescher}},
\newblock \bibinfo{title}{{The JINA REACLIB Database: Its Recent Updates and Impact on Type-I X-ray Bursts}},
\newblock \bibinfo{journal}{Astrophyical Journal Supplement Series} \bibinfo{volume}{189} (\bibinfo{year}{2010}) \bibinfo{pages}{240--252}. \DOIprefix\doi{10.1088/0067-0049/189/1/240}.

\end{thebibliography}


\end{document}